\documentclass{emulateapj}

\def\lapp{\ifmmode\stackrel{<}{_{\sim}}\else$\stackrel{<}{_{\sim}}$\fi}
\def\gapp{\ifmmode\stackrel{>}{_{\sim}}\else$\stackrel{>}{_{\sim}}$\fi}
\def\dmunit{$\mathrm{pc\,cm^{-3}}$\xspace}

\usepackage{xspace}
\usepackage[usenames, dvipsnames]{color}
\usepackage{soul}
\usepackage{enumerate}
\usepackage{amsmath}
\usepackage{float}
\usepackage[caption = false]{subfig}
\usepackage{hyperref}

\newcommand{\presto}{\texttt{PRESTO}\xspace}

\newcommand{\prepfold}{\texttt{prepfold}\xspace}
\newcommand{\spsearch}{\texttt{single\_pulse\_search.py}\xspace}

\newcommand{\tempotwo}{\texttt{TEMPO2}\xspace}

\slugcomment{Accepted version August 3, 2015} % change automatic 'draft' text
\begin{document}

\title{Discovery and Follow-up of Rotating Radio Transients with the Green Bank and LOFAR Telescopes}

\author{
C.~Karako-Argaman\altaffilmark{1}, V.~M.~Kaspi\altaffilmark{1}, R.~S.~Lynch\altaffilmark{2,3}, J.~W.~T.~Hessels\altaffilmark{4,5}, V.~I.~Kondratiev\altaffilmark{4,6}, M.~A.~McLaughlin\altaffilmark{3}, S.~M.~Ransom\altaffilmark{7}, A.~M.~Archibald\altaffilmark{4}, J.~Boyles\altaffilmark{8}, F.~A.~Jenet\altaffilmark{9}, D.~L.~Kaplan\altaffilmark{10,11}, L.~Levin\altaffilmark{3}, D.~R.~Lorimer\altaffilmark{3,17}, E.~C.~Madsen\altaffilmark{1}, M.~S.~E.~Roberts\altaffilmark{12,13}, X.~Siemens\altaffilmark{10}, I.~H.~Stairs\altaffilmark{14}, K.~Stovall\altaffilmark{9,15,16}, J.~K.~Swiggum\altaffilmark{3}, J.~van Leeuwen\altaffilmark{4,5}}

\altaffiltext{1}{Department of Physics, McGill University, 3600
  University Street, Montreal, QC H3A 2T8, Canada {\tt karakoc@physics.mcgill.ca}}
\altaffiltext{2}{National Radio Astronomy Observatory, P.O.\ Box 2,
  Green Bank, WV 24944, USA}
\altaffiltext{3}{Department of Physics, West Virginia University, 111
  White Hall, Morgantown, WV 26506, USA}
\altaffiltext{4}{ASTRON, the Netherlands Institute for Radio Astronomy,
  Postbus 2, 7990 AA Dwingeloo, The Netherlands}
\altaffiltext{5}{Anton Pannekoek Institute for Astronomy, University
  of Amsterdam, Science Park 904, 1098 XH Amsterdam, The Netherlands}
\altaffiltext{6}{Astro Space Center of the Lebedev Physical Institute,
  Profsoyuznaya str. 84/32, Moscow 117997, Russia}
\altaffiltext{7}{National Radio Astronomy Observatory, 520 Edgemont
  Road, Charlottesville, VA 22903, USA}
\altaffiltext{8}{Department of Physics and Astronomy, Western Kentucky
  University, Bowling Green, KY 42101, USA}
\altaffiltext{9}{Center for Advanced Radio Astronomy and Department
  of Physics and Astronomy, University of Texas at Brownsville,
  Brownsville, TX 78520, USA}
\altaffiltext{10}{Department of Physics, University of Wisconsin, 
  Milwaukee, Milwaukee WI 53211 USA}
\altaffiltext{11}{Department of Astronomy, University of Wisconsin-Madison, 
  475 North Charter Street, Madison, Wisconsin 53706-1582, USA}
\altaffiltext{12}{Eureka Scientific Inc., 2452 Delmer Street, Suite
  100, Oakland, CA 94602, USA}
\altaffiltext{13}{Department of Physics, Ithaca College, Ithaca, NY
  14850, USA}
\altaffiltext{14}{Department of Physics and Astronomy, University of
  British Columbia, 6224 Agricultural Road, Vancouver, BC V6T 1Z1,
  Canada}
\altaffiltext{15}{Department of Physics and Astronomy, University of
  Texas at San Antonio, San Antonio, TX 78249, USA}
\altaffiltext{16}{Department of Physics and Astronomy, University of 
  New Mexico, Albuquerque, NM 87131, USA}
\altaffiltext{17}{Adjunct Astronomer at the National Radio Astronomy Observatory}

\begin{abstract}
We have discovered 21 Rotating Radio Transients (RRATs) in data from the Green Bank Telescope (GBT) 350-MHz Drift-scan and the Green Bank North Celestial Cap pulsar surveys using a new candidate sifting algorithm. RRATs are pulsars with sporadic emission that are detected through their bright single pulses rather than Fourier domain searches. We have developed {\tt RRATtrap}, a single-pulse sifting algorithm that can be integrated into pulsar survey data analysis pipelines in order to find RRATs and Fast Radio Bursts. We have conducted follow-up observations of our newly discovered sources at several radio frequencies using the GBT and Low Frequency Array (LOFAR), yielding improved positions and measurements of their periods, dispersion measures, and burst rates, as well as phase-coherent timing solutions for four of them. 
The new RRATs have dispersion measures (DMs) ranging from 15 to 97~\dmunit, periods of 240~ms to 3.4~s, and estimated burst rates of 20 to 400~pulses~hr$^{-1}$ at 350~MHz. 
We use this new sample of RRATs to perform statistical comparisons between RRATs and canonical pulsars in order to shed light on the relationship between the two populations. We find that the DM and spatial distributions of the RRATs agree with those of the pulsars found in the same survey. We find evidence that slower pulsars (i.e. $P>200$~ms) are preferentially more likely to emit bright single pulses than are faster pulsars ($P<200$~ms), although this conclusion is tentative. 
Our results are consistent with the proposed link between RRATs, transient pulsars, and canonical pulsars as sources in various parts of the pulse activity spectrum.
\end{abstract}


\section{Introduction}
\label{sec:Introduction}
Rotating Radio Transients (RRATs) are a recently discovered class of radio-emitting neutron stars \citep{mll+06}, which are characterized by sporadic emission of individually detectable single pulses. Unlike regular pulsars, RRATs typically emit a bright pulse, then have no detectable emission for many rotations, before another such pulse is emitted. The known RRAT population is still very small, with only $\sim$100 RRATs known\footnote{See the `RRATalog' for details on all known RRATs: \url{http://astro.phys.wvu.edu/rratalog}}, of which only $\sim$25 have phase-connected timing solutions, making the characterization of this source class difficult. 

The reason for RRATs' irregular emission is not yet understood, although various models exist. 
One proposed explanation by \cite{wsr+06} is that RRATs are simply faint pulsars with high pulse-to-pulse amplitude variability, such that only the few brightest pulses from these sources are detected. Another suggestion is that RRAT emission is extrinsically modulated, for example by an asteroid belt \citep{cs08,li06}. \cite{zgd07} alternatively propose that RRATs may be old neutron stars whose emission is starting to turn off, or that they are related to nulling pulsars. Specifically, \cite{bsb10} suggest that RRATs are an extreme case of nulling pulsars. It is also possible that RRATs are explained by some combination of the above scenarios.

Nulling pulsars are intermittent sources that typically emit several consecutive pulses, followed by a short period of no pulses, until the next set of pulses is detected. In this case, most RRATs would be pulsars with nulling fractions of $> 99\%$. It may also be that the nulling fraction of pulsars increases over their lifetime \citep[e.g.][]{rit76,wmj07}. 
RRATs would thus represent a late evolutionary phase in a pulsar's life according to this hypothesis. 
This relation between RRATs and nulling pulsars is supported by the discovery of a source that switches modes between a bright pulsar with a low nulling fraction and a classic RRAT \citep{bsb10}.

However, the link between RRATs, canonical pulsars, and other apparent pulsar subclasses is not yet clear. 
It has been proposed that RRATs, some of which have exhibited long periods and magnetic fields somewhat higher than the average slow pulsar field, may be related to magnetars and X-ray dim isolated neutron stars (XDINS) \citep[e.g.][]{mll+06,lmk+09,ptp06}. 

Regardless of whether RRATs are a physically distinct class of pulsars, 
or just at the tail of the distribution in terms of pulse-to-pulse variability, 
the enormous bias that exists against finding them in pulsar surveys argues that they may represent a large fraction of the radio-detectable neutron star (NS) population, perhaps even outnumbering regular radio pulsars \citep{mll+06}.  
Moreover, \cite{kk08} find that when the radio pulsar population is corrected to include the potentially large RRAT contribution, the Galactic core-collapse supernova rate is too low to explain the NS birth rate. 
This problem is solved by those models that state that RRATs are evolutionarily linked to other pulsar classes. 
If, on the other hand, RRATs do represent a distinct class of sources, this problem remains unsolved. 
Finding and studying more RRATs is therefore an important step in understanding their nature, and better understanding the pulsar population as a whole.

Along with RRATs, over the last few years there have been discoveries of other fast (sub-second timescales) transients, characterized by isolated bright dispersed single pulses. The first of these was the ``Lorimer burst" \citep{lbm+07}, a pulse with a seemingly extragalactic origin due to its high dispersion measure (DM), though its extragalactic origin has been questioned \citep[e.g.][]{kon+14}. Another phenomenon is ``perytons," which are short frequency-swept pulses at high DMs, but are thought to be associated with terrestrial atmospheric events \citep{bbe+11}. \cite{kle+10} reported another high-DM burst, though its DM excess was not sufficiently large to determine whether it was cosmological or a burst from a distant Galactic RRAT. Finally, in the last two years, a phenomenon known as Fast Radio Bursts (FRBs) emerged, which are bright, high-DM pulses that are plausibly extragalactic and associated with a cataclysmic event based on their non-repeating nature \citep[see, e.g.,][]{tsb+13,sch+14}. Moreover, all of the fast transients described above have only been found at a frequency of $\sim1.4$~GHz so far. 
Characterizing new transient, highly dispersed millisecond radio bursts found in single-pulse searches is therefore important.

Here we report on a search for RRATs in two radio pulsar surveys conducted at 350~MHz, the Green Bank Telescope Drift-scan survey (hereafter `Drift-scan') and the Green Bank North Celestial Cap (GBNCC) survey, using a new search technique. This resulted in the discovery of 21 new RRATs. We have followed up these RRATs using the Green Bank Telescope (GBT) and Low Frequency Array (LOFAR) telescopes, resulting in improved positions, periods and DMs. We have also obtained phase-coherent timing solutions for four of the RRATs.

We describe the two surveys in Section~\ref{sec:surveys}, the data analysis, {\tt RRATtrap} candidate sifting algorithm\footnote{{\tt RRATtrap} is available at \url{http://github.com/ckarako/rrattrap}}, and discoveries in Section~\ref{sec:dataanalysis}, and the follow-up analysis in Section~\ref{sec:followup}. We discuss our results in the context of the general pulsar population in Section~\ref{sec:discussion}, and conclude in Section~\ref{sec:conclusion}.

\section{Survey observations}
\label{sec:surveys}
The Drift-scan and GBNCC surveys were both conducted using the Robert C. Byrd Green Bank Telescope, whose 100-m diameter, unblocked aperture, and low-noise receivers make it an extremely sensitive and powerful tool for radio astronomy. Here we describe these surveys, including their sky coverage, sensitivities, and pulsar discoveries. We summarize observational parameters for the two surveys in Table~\ref{table:obsmodes}. More details on the Drift-scan are presented in \cite{blr+13} and \cite{lbr+13}, and on the GBNCC in \cite{slr+14}.

\subsection{Green Bank Telescope Drift-scan survey}
\label{subsec:surveys drift}
\begin{figure*}
    \centering
    \includegraphics[width=0.85\linewidth]{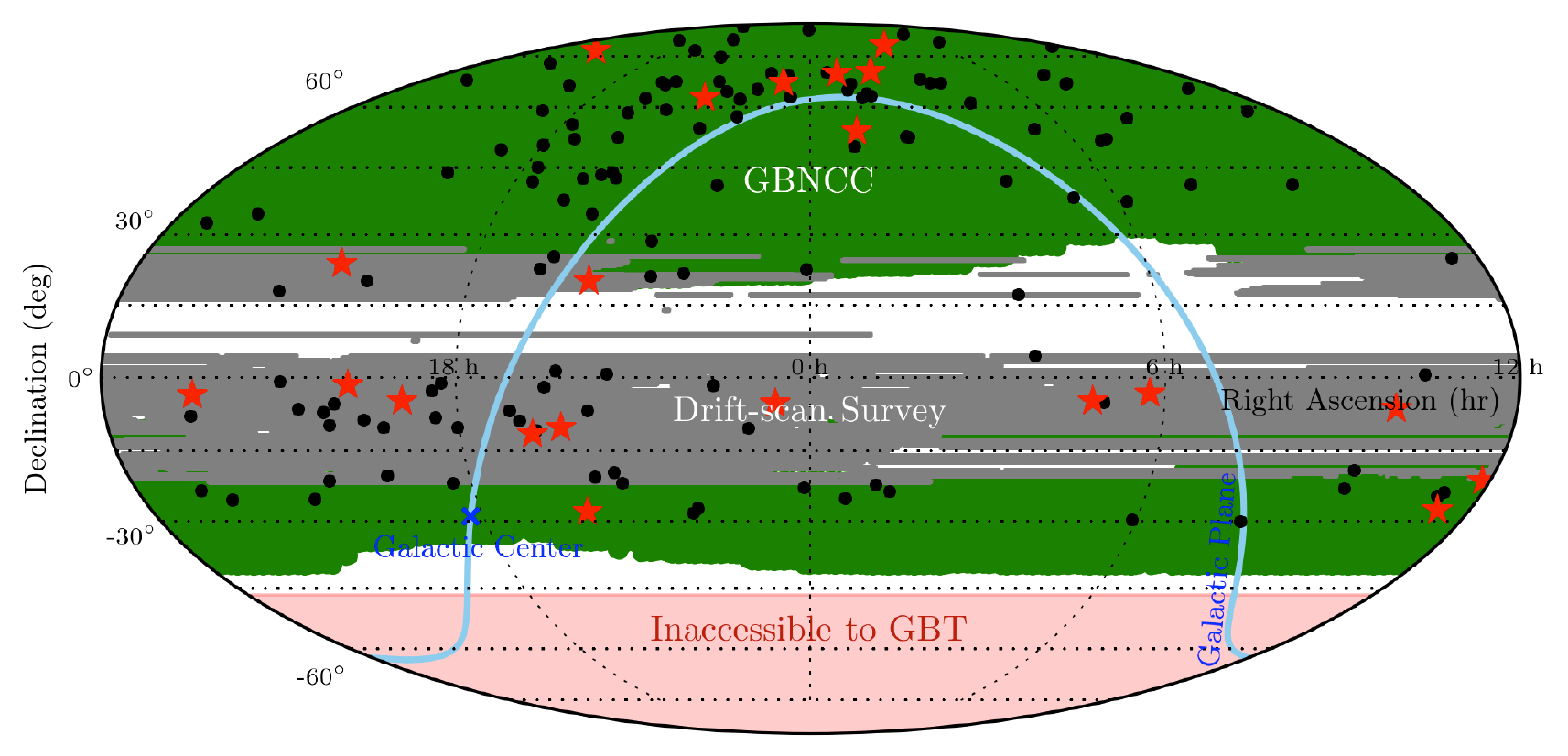}
    \caption{Sky coverage of the GBT Drift-scan and GBNCC surveys, plotted in Equatorial coordinates. The gray strips indicate the areas of sky observed in the Drift-scan, while in green are the areas observed in the GBNCC survey. Discoveries from each survey are also indicated: pulsars are denoted by black circles, and RRATs are denoted by red stars. The sky coverage and discoveries of the GBNCC are current up to 2015 March.\label{fig:skycoverage}} %plot current up until Mar 3, 2015. Includes 10 GBNCC RRATs, 11 drift RRATs, 93 GBNCC psrs (excl RRATs)
\end{figure*}

The GBT Drift-scan survey was a radio pulsar survey conducted during the northern summer of 2007, while the azimuth track of the telescope was undergoing repairs. Although the GBT is normally fully steerable, this survey took advantage of the time otherwise spent idle by pointing the telescope at several fixed elevations, and collecting data as the sky drifted over it. 
Overall, this survey produced 134 TB of data over 1491 observing hours, and covered about 10300 square degrees of the sky. 
The coverage of the Drift-scan spanned declination ($\delta$) ranges of $-8^{\circ} \lesssim \delta \lesssim +38^{\circ}$ and $-21^{\circ} \lesssim \delta \lesssim +38^{\circ}$, depending on the telescope's azimuth (which varied between two values), and is shown in Figure~\ref{fig:skycoverage}. %azimuth vals: 229 deg and 192 deg, resp.

Although the recording of data was continuous for each observing session, the data were divided into pseudo-pointings, each 140~s in duration, roughly corresponding to the time it takes a point on the sky to pass through the full-width half-maximum (FWHM) of the telescope beam. This resulted in about 30000 pseudo-pointings, each independently analyzed as described in Section~\ref{sec:dataanalysis}. 
Each pointing was searched up to a DM of 1015~\dmunit, with $\sim$10000 DM trials. The {\tt RRATtrap} algorithm was run on the single-pulse results from all beams, constituting $1491~\mathrm{hr} \times 0.28~\mathrm{deg}^2 = 418~\mathrm{hr~deg}^2$ of sky. 
% beam area = pi*r^2, r = FWHM/2, FWHM = 1.22*lambda/D = 1.22*(c/0.35 GHz) rads = 0.6 deg --> beam area = pi*(0.6 deg/2)^2 = 0.28 deg^2 
Note that the Drift-scan analysis was conducted before the FRB phenomenon was unambiguously confirmed \citep{tsb+13}, and thus the data were not searched to such high DMs as for the GBNCC (see Section~\ref{subsec:surveys gbncc}). A reprocessing of the data with a search of higher DMs is now beginning.

This survey yielded 35 new pulsars.  
Amongst the new pulsars are some exotic systems, including a millisecond pulsar in a hierarchical triple system with two white dwarf companions \citep[PSR J0337+1715;][]{rsa+14}, and a radio pulsar/low mass X-ray binary transition object \citep[PSR J1023+0038;][]{asr+09}.  
Furthermore, applying the new single-pulse sifting techniques developed here, we have discovered 11 RRATs, and 5 as-yet-unconfirmed RRAT candidates. These RRAT discoveries are presented in Table~\ref{table:rratdiscoveries} and elaborated upon in Sections~\ref{subsec:results} and \ref{subsec:timedrrats}. The RRAT candidates, as well as the number and duration of re-observations, are presented in Table~\ref{table:rratcands}.

\subsection{Green Bank North Celestial Cap survey} 
\label{subsec:surveys gbncc}
The GBNCC survey is an ongoing pulsar survey that began in 2009 and which also uses the GBT at 350~MHz. Unlike the Drift-scan, this survey was designed with the goal of uniformly covering the sky visible to the GBT. The survey makes use of the Green Bank Ultimate Pulsar Processing Instrument \citep[GUPPI;][]{drd+08}, a powerful backend that is now routinely used for pulsar observations at the GBT.

The overall region to be observed is divided into pointings, and over the search campaign, each pointing is observed for 120~s. 
The region that will be covered at the completion of the survey is the entire GBT-visible sky, $-45^{\circ} < \delta < 90^{\circ}$, and consists of about 125000 pointings. Figure~\ref{fig:skycoverage} shows the current sky coverage of the survey. So far, over 66000 pointings have been observed, representing about half of the total planned survey area. %66000+ current up until March 3, 2015 
In the early processing, each pointing was searched up to a DM of 500~\dmunit, with $\sim$17500 DM trials, but in the last year the maximum DM has been increased to 3000~\dmunit in order to search for highly-dispersed, possibly extragalactic bursts. The previously analyzed pointings have not yet been reprocessed with this increased maximum DM. 
This DM threshold was chosen to be substantially larger than the highest observed DM of FRBs, $\sim$1100~\dmunit \citep{tsb+13}, and was limited to 3000~\dmunit, since at this DM the time delay across the GBNCC observing band is $\sim$60~s, which is half of the observation duration. 
Approximately 200 hr deg$^2$ have so far been analyzed with {\tt RRATtrap}. %120s/beam * 21300 beams * 0.28 sq deg = 198.8 hr deg^2; 21300 approximation of gbncc beams processed

It is estimated that the GBNCC survey will be about 2.5 times more sensitive to low-DM pulsars at high Galactic latitudes compared to past and some ongoing pulsar surveys of the same region \citep{slr+14}. The data analysis for this survey consists of the techniques described in Section~\ref{sec:dataanalysis}, along with improved methods of examining candidates using an online database.

The survey has so far been successful in finding 93 %correct as of March 3, 2015
new pulsars, as well as 10 new RRATs and 1 RRAT candidate which were discovered using the algorithm described in this work.  
Among the newly discovered pulsars are 4 millisecond pulsars which have been included in pulsar timing arrays for the purpose of detecting gravitational waves \citep[e.g.][]{haa+10}, a double neutron star system (Lynch et al., in prep), and a binary system with an unusual degenerate companion \citep[PSR J1816+4510;][]{ksr+12}.

\subsection{Survey sensitivity}
\label{subsec:surveys sensitivity}
The sensitivities of the Drift-scan and the GBNCC surveys to regular pulsars are presented by \cite{lbr+13} and \cite{slr+14}, respectively. Here we compute these surveys' sensitivities to single-pulse sources, following \cite{cm03}:
\begin{equation}
\label{eq:spsensitivity}
S_i = \frac{\beta (\mathrm{S/N})_b\, (T_{\mathrm{sys}}+T_\mathrm{sky})}{G\, W_i}\sqrt{\frac{W_b}{n_{\mathrm{p}}\Delta f}}\ ,
\end{equation}
where $S_i$ is the intrinsic flux density of the pulse, $\beta$ is a factor accounting for sensitivity losses due to digitization, (S/N)$_b$ is the measured signal-to-noise ratio of the broadened pulse, $T_{\mathrm{sys}}$ is the system temperature at the frequency of observation, $T_\mathrm{sky}$ is the sky temperature, $G$ is the telescope gain, $W_i$ and $W_b$ are the intrinsic and broadened pulse widths, respectively, $n_{\mathrm{p}}$ is the number of summed polarizations, and $\Delta f$ is the bandwidth.

The detected pulse width, $W_b$, depends not only on the intrinsic pulse width, but also on ISM propagation effects such as dispersion and scattering, which smear the pulse. Although dedispersion and a large number of frequency channels are used in order to mitigate pulse smearing due to dispersion, it is impossible to correct for the dispersion smearing across an individual frequency channel. The broadened pulse width is given by
\begin{equation}
\label{eq:wb}
W_b = \sqrt{W_i^2 + t_\mathrm{samp}^2 + t_\mathrm{chan}^2 + t_\mathrm{scatt}^2}\ ,
\end{equation}
where  
$t_\mathrm{samp}$ is the sampling time, $t_\mathrm{chan}$ is the dispersive smearing within each frequency channel, and $t_\mathrm{scatt}$ is the scatter broadening time related to the effect of multi-path scattering of signals due to irregularities in the ISM \citep[e.g.][]{ric90}. 
An expression for $t_\mathrm{chan}$ can be found in \cite{lk04}; scaling to Drift-scan/GBNCC observing frequency and channel bandwidth, 
\begin{equation}
\label{eq:tchan}
t_\mathrm{chan} = 4.7~\mu s \times \left(\frac{\Delta f_\mathrm{chan}}{0.0244~\mathrm{MHz}}\right) \times \left(\frac{f}{350~\mathrm{MHz}}\right)^{-3} \times\left(\frac{\mathrm{DM}}{\mathrm{pc\, cm}^{-3}}\right)\ .
\end{equation}
An approximation for $t_\mathrm{scatt}$ is given by \cite{cor02},
\begin{multline} % split line, for 2 column format
\label{eq:tscatt}
% split line, for 2 columns:
\log\left(\frac{t_\mathrm{scatt}}{\mu s}\right) = -3.59 + 0.129\log\mathrm{DM} \\+ 1.02(\log\mathrm{DM})^2-4.4\log\left(\frac{f}{\mathrm{GHz}}\right)\ ,
\end{multline}
with $\Delta f_\mathrm{chan}$ the channel bandwidth and $f$ the central observing frequency.

\begin{figure}
\centering
\includegraphics[width=\linewidth]{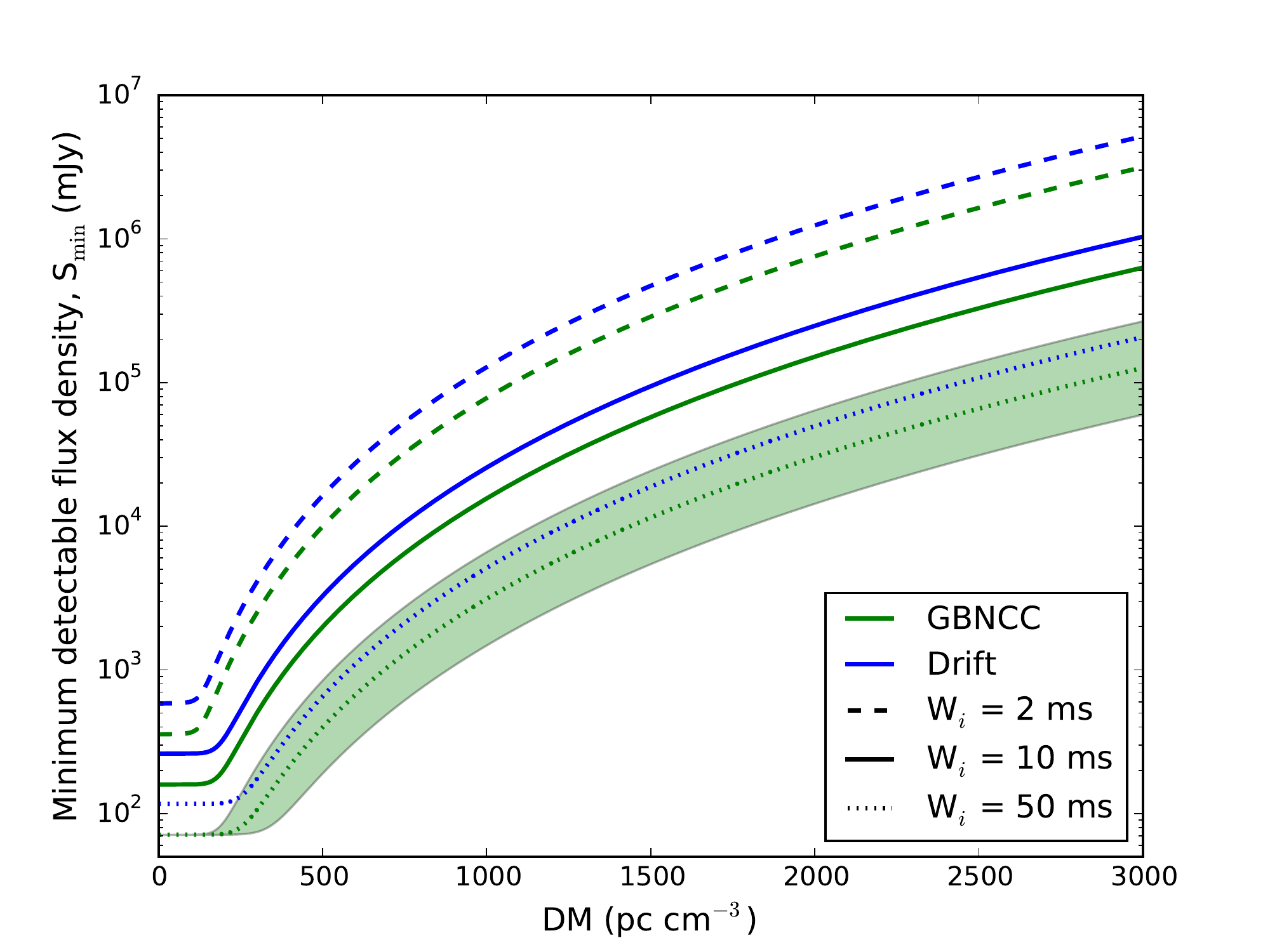}
\caption{Minimum detectable flux density, $S_\mathrm{min}$, of single pulses as a function of dispersion measure, for intrinsic pulse widths $W_i$ of 2, 10, and 50~ms, plotted for the Drift-scan (blue) and GBNCC (green) surveys. The green band shows the scatter due to the variance in measured pulse broadening (Equation~\ref{eq:tscatt}) for the $W_i=50$~ms GBNCC curve, with similar bands applicable for the other curves. \label{fig:spsensitivity}}
\end{figure}

For a minimum detected signal-to-noise ratio (S/N)$_b$ of 6 and an assumed intrinsic pulse width, we can compute each survey's sensitivity to single-pulse sources by substituting the survey-specific parameters from Table~\ref{table:obsmodes} into Equation~\ref{eq:spsensitivity}. We plot the resulting sensitivity as a function of DM in Figure~\ref{fig:spsensitivity}, for intrinsic pulse widths of 2~ms, 10~ms (the typical width of the RRAT pulses we detected), and 50~ms, for the Drift-scan and GBNCC surveys. 

\begin{deluxetable*}{lccccc} 
    \tablecaption{Observing modes used in RRAT discovery and follow-up observations.\label{table:obsmodes}}
    \tablewidth{0pt}
    \tablehead{ Parameters & \colhead{GBT Drift-scan} & \colhead{GBNCC} & \colhead{GBT 350} & \colhead{GBT 820} & \colhead{LOFAR} \\
                &                          &                 & \colhead{follow-up}& \colhead{follow-up}& }
    \startdata
    Center freq. (MHz)      & 350           & 350   & 350      & 820       & 150  \\
    Bandwidth (MHz)         & 50            & 100   & 100      & 200       & 80   \\
    Number of freq. channels& 2048          & 4096  & 2048     & 2048      & 6400 \\
    Sampling time ($\mu$s)  & 81.92         & 81.92 & 81.92    & 81.92     & 655.36\\
    Obs. duration (min)     & 2.3           & 2     & 10--15   & 6--15     & 10--30\\
    Beam FWHM (arcmin)      & 36            & 36    & 36       & 15        & 5    \\
    Backend                 & Spigot        & GUPPI & GUPPI    & GUPPI     & Blue Gene/P \\
    Analog-to-digital conversion factor, $\beta$ & 1.16 & 1 & 1 & 1 & 1 \\
    %System temperature (K)  & 75            & 75    & 75       & 38        & 900\tablenotemark{a}  \\ % includes average Tsky's
    System temperature (K)\tablenotemark{a}  & 23            & 23    & 23       & 23        & 400\tablenotemark{b}  \\ % does not include Tsky
    Gain (K/Jy)             & 2             & 2     & 2        & 2         & 5    \\
    Sensitivity limit (mJy) & 260           & 160   & 160      & 45        & 80\tablenotemark{c}   \\
    \enddata
    \tablecomments{The sensitivity limit quoted above is for single-pulse sources, and is computed as in Section~\ref{subsec:surveys sensitivity}, for a DM of 100~\dmunit. \tablenotetext{a}{The system temperature we report does not include the sky temperature.} \tablenotetext{b}{The average receiver temperature is reported by \cite{akw13}.}\tablenotetext{c}{The sensitivity for LOFAR is obtained from \cite{sha+11} after scaling to the full available bandwidth, and assuming a 10-ms pulse width as in the calculations for the GBT observations.}}
\end{deluxetable*}

\section{RRAT search algorithm}
\label{sec:dataanalysis}
The data from the Drift-scan and GBNCC surveys were analyzed using standard tools from the \presto suite of pulsar search software \citep{ran01}\footnote{\url{http://www.cv.nrao.edu/~sransom/presto}}. This analysis includes RFI removal, dedispersion, single-pulse searches, and periodicity searches, as described in \cite{lbr+13} for the Drift-scan and \cite{slr+14} for GBNCC.
{\tt RRATtrap}, the single-pulse sifting algorithm developed here, was then applied to the outputs of the \presto single-pulse search routine in order to search for RRATs. 
We describe our algorithm and present its results for the Drift-scan and GBNCC surveys.

\subsection{Single-pulse sifting}
\label{sec:spsifting}
The data are searched for bright single pulses using \presto's \spsearch. This tool convolves the dedispersed and downsampled time series with box-car filters of various widths (namely 1, 2, 3, 4, 6, 9, 14, 20, 30, 45, 70, 100, and 150 time bins) and records events which exceed the mean by an imposed threshold, here 5 standard deviations \citep[see][for details on matched filtering]{cm03}. \spsearch does not suffer from the `root 2 problem' reported by \citet{kp15}, in which the sensitivity to pulses varies depending on the alignment of events with time series samples. This is an issue for programs which use power-of-two downsampling during the single-pulse search, rather than downsampling the data before searching for pulses and using matched filtering.

A list and 4--5 plots of all $\mathrm{S/N} \geq 5$ candidate single-pulse events in the beam are then saved, with each plot covering a portion of the DM range searched.
Since single-pulse plots are produced for every pointing, the number of resulting plots for a full survey is very large (e.g. $4\times30000$ for the Drift-scan, and $5\times66000$ for GBNCC so far), making their examination a tedious task with potentially subjective outcomes. We have developed an automated sifting algorithm in order to identify RRAT candidates and astrophysical pulses in the output of single-pulse searches, eliminating the need for manual inspection of each diagnostic plot produced. This algorithm is described below and illustrated in Figure~\ref{fig:algflowchart}. An example output plot from the algorithm is shown in Figure~\ref{fig:codespplot}.

\subsubsection{Algorithm design}
\label{subsubsec:algorithm design}
The single-pulse sifting algorithm is designed to take advantage of properties that characterize astrophysical signals, and use these to distinguish astrophysical signals from other signals. Our algorithm is based on the following concepts:
\begin{enumerate}[(a)]
  \item{\label{dmrange} A bright signal will be detected over a range of DMs, with the strongest detection at the optimal DM and weaker detections above and below this DM due to smearing of the pulse when it is dedispersed at incorrect DMs. The expected S/N fall-off with incorrect DM is given by \cite{cm03}, Equation~12.} 
  \item{\label{rfidm0} Since signals are strongest at the optimal DM, we expect that signals of terrestrial origin (namely RFI) will peak at a DM of 0~\dmunit. We can thus classify any signals that peak at $\mathrm{DM}\sim0$~\dmunit as not astrophysical and reject them.}
  \item{\label{rfispan} RFI that is only present in a narrow range of frequencies will not be subject to dispersion effects, and thus will appear consistently bright over a very large range of DMs.} 
\end{enumerate}
Concept (\ref{dmrange}) means that a given pulse, whether astrophysical or not, will be associated with many statistically significant ``single-pulse events" that will be found in the single-pulse search. These events will be spread over a small range of DMs, and will occur at approximately the same time. The first step in our algorithm is thus to group events that belong to the same pulse by checking whether they satisfy this criterion, that is, lie within some small window of DM and time. Once all single-pulse events in a beam have been divided into groups, we examine each group's collective properties in order to decide whether it behaves like an astrophysical pulse, and rate it based on these results.

The first test we employ is related to group size. If a group has too few ($<20$, a tunable parameter) events, we classify it as noise, and assign it a rank of 1. 
We found that this minimum group size was sufficient in order to remove much of the noise, while still maintaining sensitivity to the weakest pulses ($\mathrm{S/N}_{\mathrm{max}}=6$). Moreover, examination of detections of known pulses has shown us that this group size is a fairly conservative (i.e. easy for a real pulse to reach) criterion. 
However, a potential future improvement might be to relate the minimum group size for each pulse to its width and peak S/N, since these affect the number of events that will be associated with a given pulse.

Next, we examine the S/N vs.~DM behavior of the remaining groups. From concept (\ref{rfidm0}), we expect that a group of events that is caused by RFI will have a peak S/N at $\mathrm{DM}\sim0$~\dmunit. Thus, any group that satisfies this criterion, namely $\mathrm{DM}(\mathrm{S/N}_{\mathrm{max}})< \mathrm{DM}_{\mathrm{min}}$, is classified as RFI, which we define by a rank of 2. Here we use $\mathrm{DM}_\mathrm{min} = 2$~\dmunit. 

We then again make use of concept (\ref{dmrange}) by looking for groups whose S/N peaks at some non-zero DM and decreases above and below that DM. 
We divide each group into five intervals in DM (numbered 1--5 from low to high DM), with each interval containing one fifth of the group's events. We refer to the maximum S/N of the events in interval $i$ as $\mathrm{S/N}_i$. If the maximum S/N of the group occurs in one of the inner intervals ($i=2, 3, 4$), and $\mathrm{S/N}_i$ drops off monotonically on either side of the peak (e.g., $\mathrm{S/N}_1 < \mathrm{S/N}_2 < \mathrm{S/N}_3$ and $\mathrm{S/N}_3 > \mathrm{S/N}_4 > \mathrm{S/N}_5$), the group is classified as a ``good" astrophysical pulse and given a rank of 4. Moreover, if such a group's peak S/N exceeds an imposed threshold, $\mathrm{S/N}_{\mathrm{thresh}} = 8$, the group is classified as ``excellent" and given a rank of 5.

If a group only somewhat follows this behavior but deviates from the expected trend (e.g. $\mathrm{S/N}_2 < \mathrm{S/N}_3$ and $\mathrm{S/N}_3 > \mathrm{S/N}_4$, but $\mathrm{S/N}_1 > \mathrm{S/N}_2$ or $\mathrm{S/N}_5 > \mathrm{S/N}_4$), it is given a rank of 3 and called ``fair."
If instead a group has mostly constant S/N and spans a large range of DMs (in this case a range $\geq100$~\dmunit, a tunable parameter), we conclude by concept (\ref{rfispan}) that it is RFI. Finally, any group with S/N behavior that does not fall into one of these categories is labelled ``other" and given a rank of~0. 

The values we used for the tunable parameters (maximum DM span and minimum number of events per group) were chosen after testing various values on a subset of data, and were found to provide a good balance between the number of pulsars identified and the total number of candidates. These parameter values will depend on the observing frequency and RFI environment of a given pulsar survey. 
We find that on average, approximately 63\% of all pulses are categorized as RFI, 11\% as fair, 13\% as good, 0.1\% as excellent, and 13\% as other. 

For each beam, colorized single-pulse plots are produced, with colors corresponding to group ratings (see Figure~\ref{fig:codespplot}). A summary text file is also saved, containing each group's characteristics including its time, minimum and maximum DM, pulse width, peak S/N, and assigned rating. These can in principle be used to sort candidate pulses, for example by assigning the characteristics different weights or applying further decision algorithms. 

Once the beams have all undergone this sifting algorithm, those that have been flagged as having ``excellent" pulses are visually examined. For those events that indeed look astrophysical, we then generate frequency vs. time plots that show arrival times of the signal in different frequency bins throughout the band. 
Since astrophysical signals experience a frequency-dependent ($1/f^{2}$) dispersive delay while propagating through the interstellar medium, we expect to see this signature if the detected pulse is indeed astrophysical. 
We thus use these plots and our knowledge of the effects of the ISM as a final means of testing the astrophysical nature of a signal, before deciding whether it is a candidate worthy of observational follow-up.

\subsubsection{Algorithm performance}
\label{subsubsec:algorithm performance}
We can assess the performance of our algorithm by quantifying its false positive and false negative rates. Moreover, we can examine the algorithm's usefulness by comparing the number of beams that it flags as potentially interesting to the total number of beams analyzed, since the latter is the number of beams that would have to be visually inspected if the sifting algorithm were not used. 
We find that on average, the algorithm identifies about 10\% of all beams as containing ``excellent" astrophysical pulses. This means that we only have to examine 10\% of all diagnostic single-pulse plots, which is a significant improvement.

The false positive rate may be computed based on the number of beams identified as containing ``excellent" pulses which, when inspected visually, are not found to contain astrophysical pulses. This misidentification is likely due to noise and unmodelled behavior of RFI that leads to clusters of single-pulse events that satisfy the imposed criteria for astrophysical pulses. We find that this rate is approximately 90\%, meaning that of every 10 beams identified, one will contain a pulsar (either known or new). If the algorithm were not used and the diagnostic single-pulse plots for every beam were examined, we would have to look through 100 beams in order to find this one pulsar. Nonetheless, we are currently working on improving the false positive rate in order to make the algorithm more effective, and further reduce the number of diagnostic plots that must be examined.

\begin{figure*}
    \centering
    \includegraphics[width=0.7\linewidth]{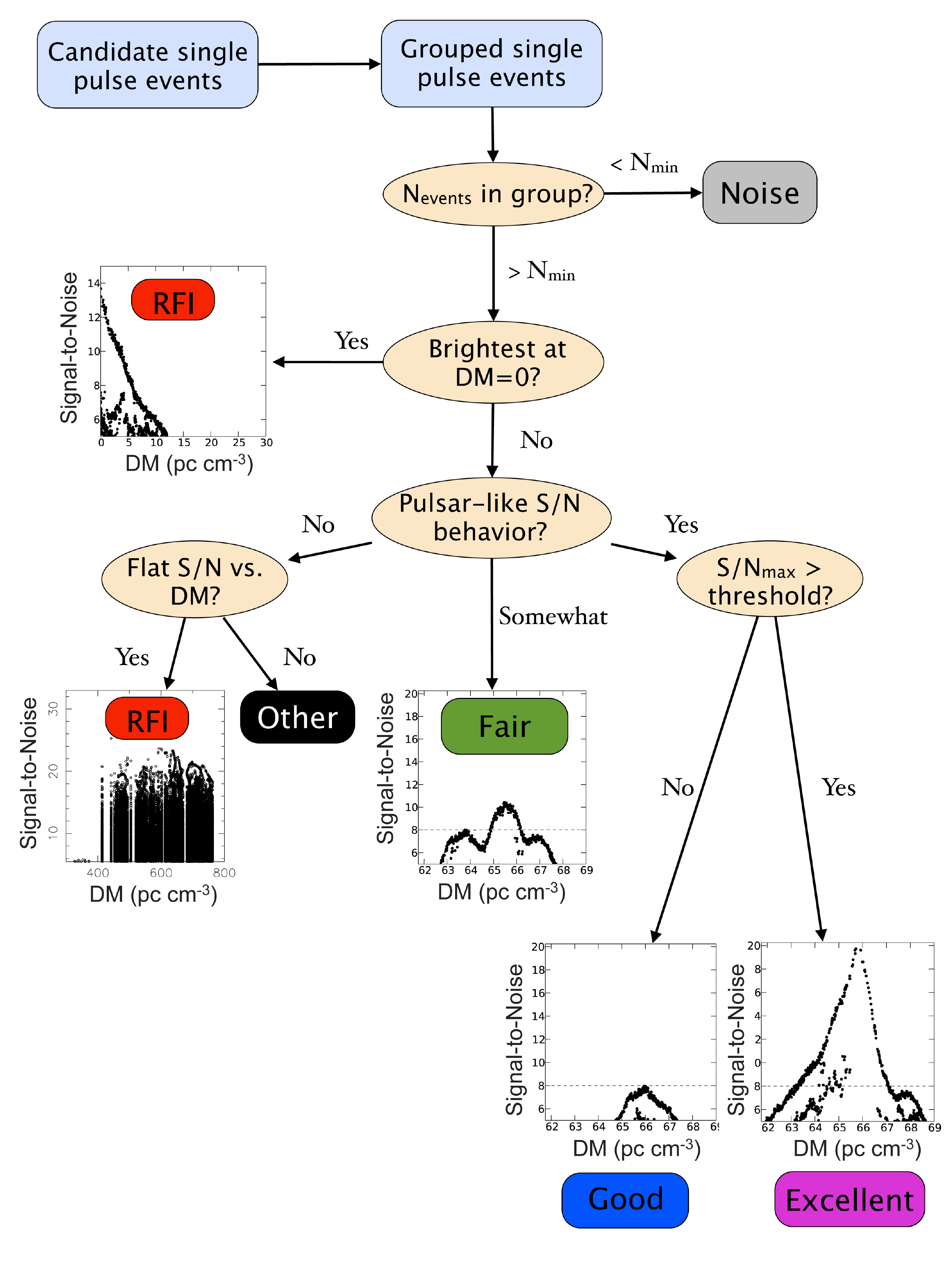}
    \caption{{\tt RRATtrap} algorithm flowchart, illustrating the rating process applied to candidate single-pulse events, as described in Section~\ref{sec:spsifting}. The threshold S/N is plotted as a dashed line in the three right plots for illustration.\label{fig:algflowchart}}
\end{figure*}

Since we use our algorithm to identify which beams should be inspected, it is also possible that some pulsars are missed. Quantifying this false negative rate is more difficult. Ideally, we would compute this rate by examining the output of the code for all beams containing known RRATs and single-pulse visible pulsars, or by performing a blind injection of synthetic RRAT signals into a sample of data files. Instead, here, we obtain a rough approximation of this rate by examining a sample of beams which were found to contain RRATs and pulsars (both new and known), and check the output of the code to see how many of their pulses were missed and not rated as ``excellent." We find that this rate is $\sim$20\%. If we loosen our criteria and count pulses which were not rated as ``excellent" or ``good," this rate decreases to $\sim$10\%, though we note that in practice only beams with top rated (``excellent") groups were examined in this work. This rate implies that there is a 20\% chance of missing an individual pulse. Note, however, that this value is an average, and in practice depends strongly on properties such as pulse shape and width, DM, and S/N. 

The false negatives in our analysis were most commonly caused by S/N~vs.~DM curves that did not conform to expectations. Qualitatively, we expect the S/N vs. DM curve to have a single maximum at the optimal DM and decrease monotonically with increasing DM differences \citep{cm03}. However, this is based on the assumption that the pulse shape has a single component. In practice, complex multi-component pulses produce asymmetric or multi-peaked S/N~vs.~DM curves (see, e.g., the `Fair' panel in Figure~\ref{fig:algflowchart}) that, if sufficiently extreme, result in false negatives. Another case in which our S/N~vs.~DM curve test may fail is for weak pulses near the noise threshold. In this case, the fluctuations in S/N due to noise are significant compared to the peak S/N of the pulse. We note, however, that both categories of false negatives described here still contained many ``fair" pulses. 
While we can make our criteria less stringent (e.g. by looking at all ``fair", ``good" and ``excellent" pulses) in order to decrease the false negative rate, doing so currently would immensely increase the false positive rate, thus lowering the effectiveness of our algorithm. However, we are working toward improving the algorithm in order to decrease the false positive and negative rates, which will also allow us to probe weaker pulse candidates.

\begin{figure}[h]
    \centering
    \includegraphics[width=\linewidth]{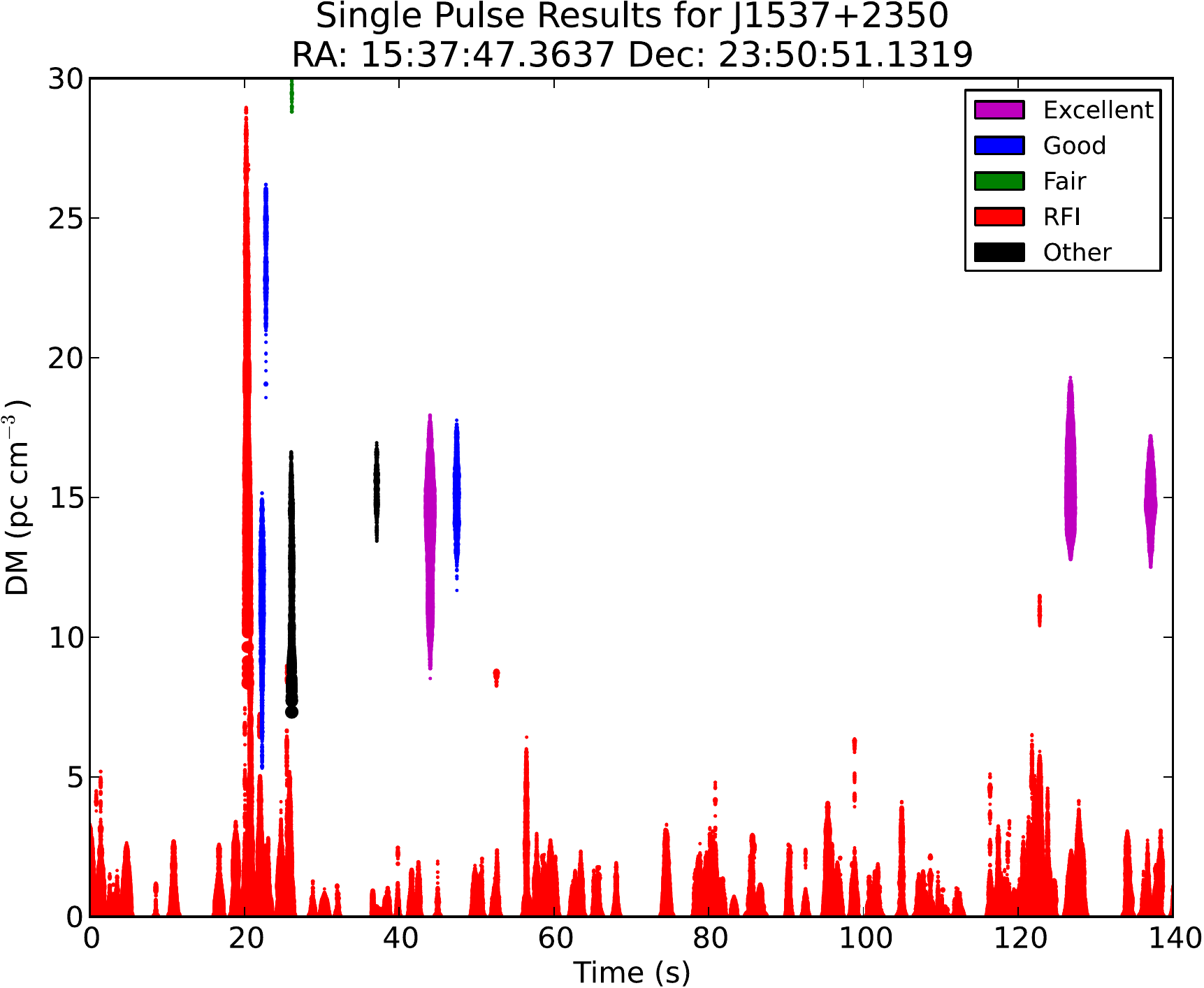}
    \caption{Example of {\tt RRATtrap} single-pulse sifting output, for newly discovered RRAT J1538+2345. Each candidate single-pulse event is plotted, with the point size proportional to the event's S/N. Radio frequency interference is identified in red, whereas 3 pulses are successfully identified in purple, and another weaker pulse in blue. These pulses all appear strongest at a common DM of $\sim$15~\dmunit. Some RFI is present but not successfully identified due to its irregular shape (seen in black and blue, at $t=20-30~\mathrm{s}$). Furthermore, a fifth pulse is visible by eye, at the same DM as the identified pulses and at $t=38~\mathrm{s}$, but is not identified by the code because its S/N does not follow the prescribed behavior (see Figure~\ref{fig:algflowchart}). 
\label{fig:codespplot}}
\end{figure}

\subsection{RRAT discoveries}
\label{subsec:results}
\begin{figure*}
    \centering
    \includegraphics[width=0.4\linewidth]{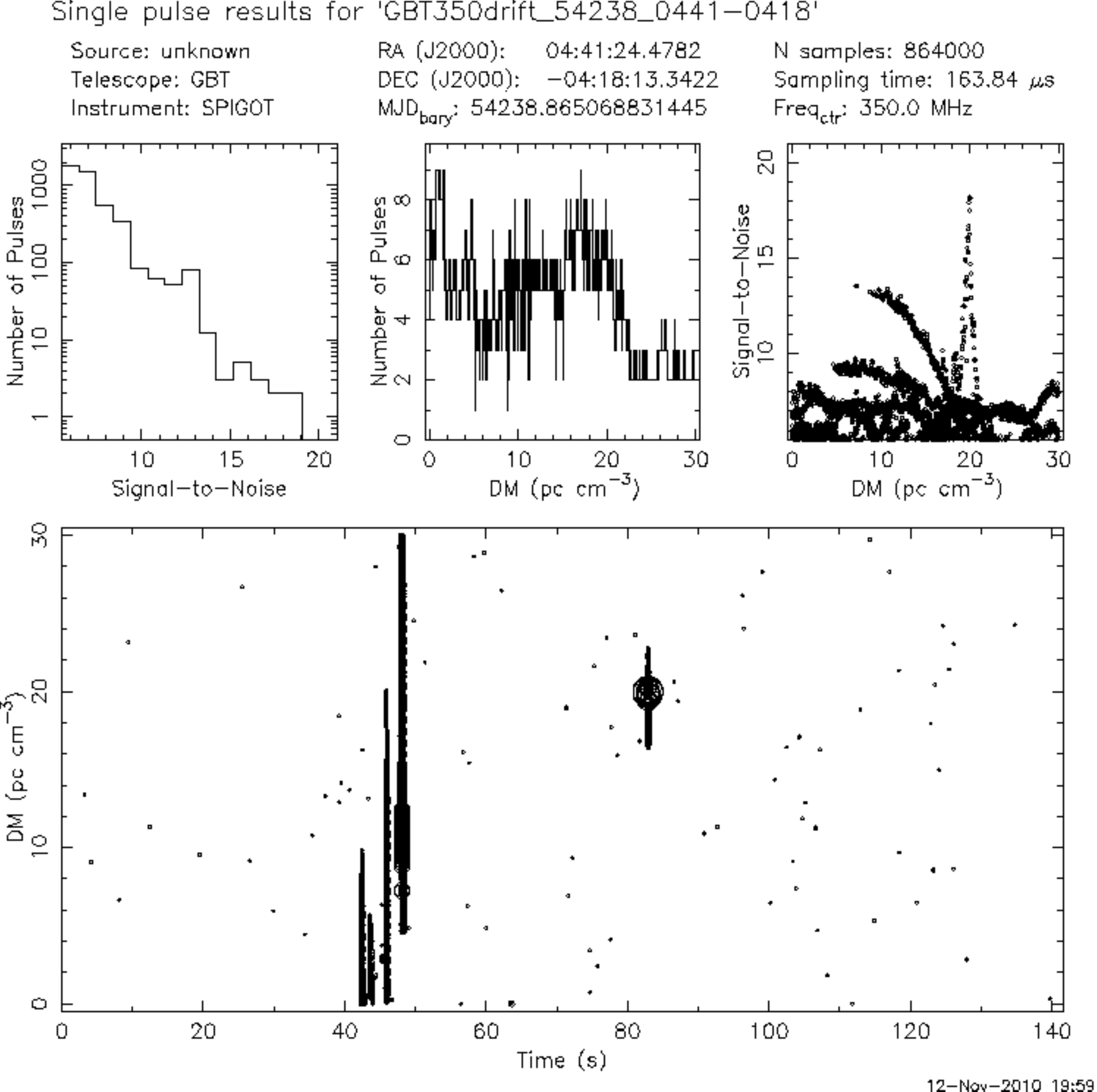}
    \includegraphics[width=0.4\linewidth]{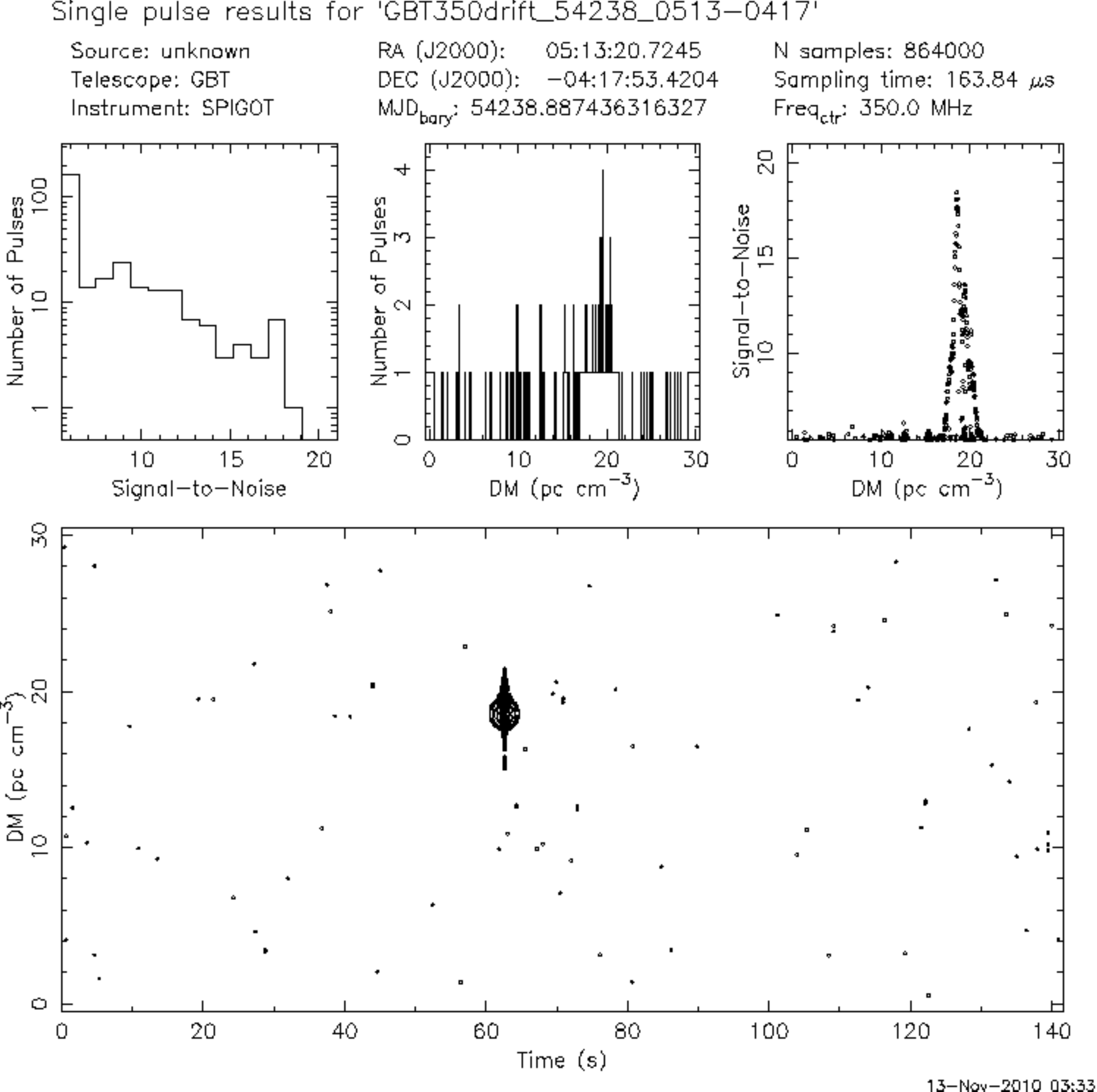}
    \\
    \vspace{0.5cm}
    \includegraphics[width=0.4\linewidth]{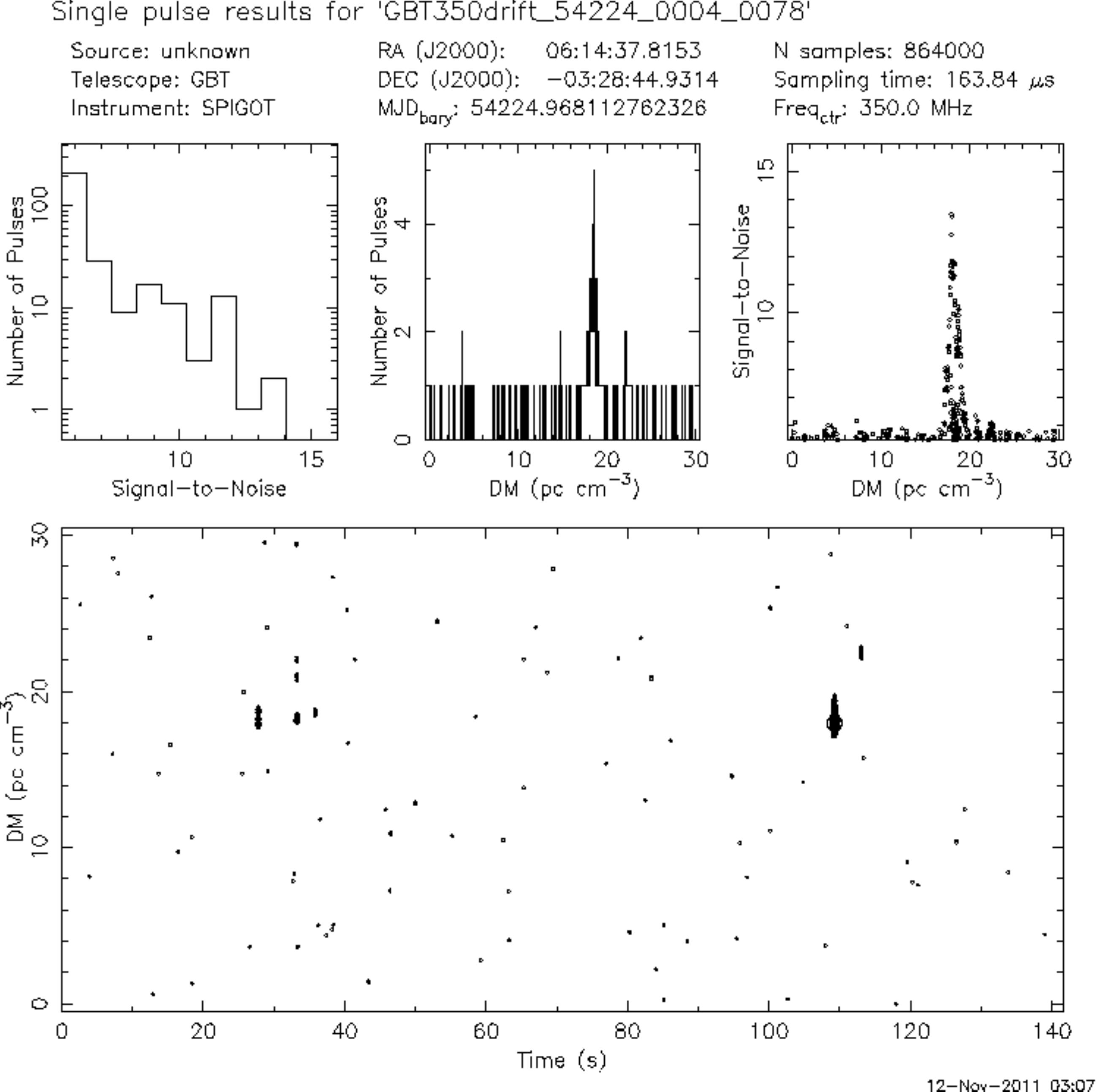}
    \includegraphics[width=0.4\linewidth]{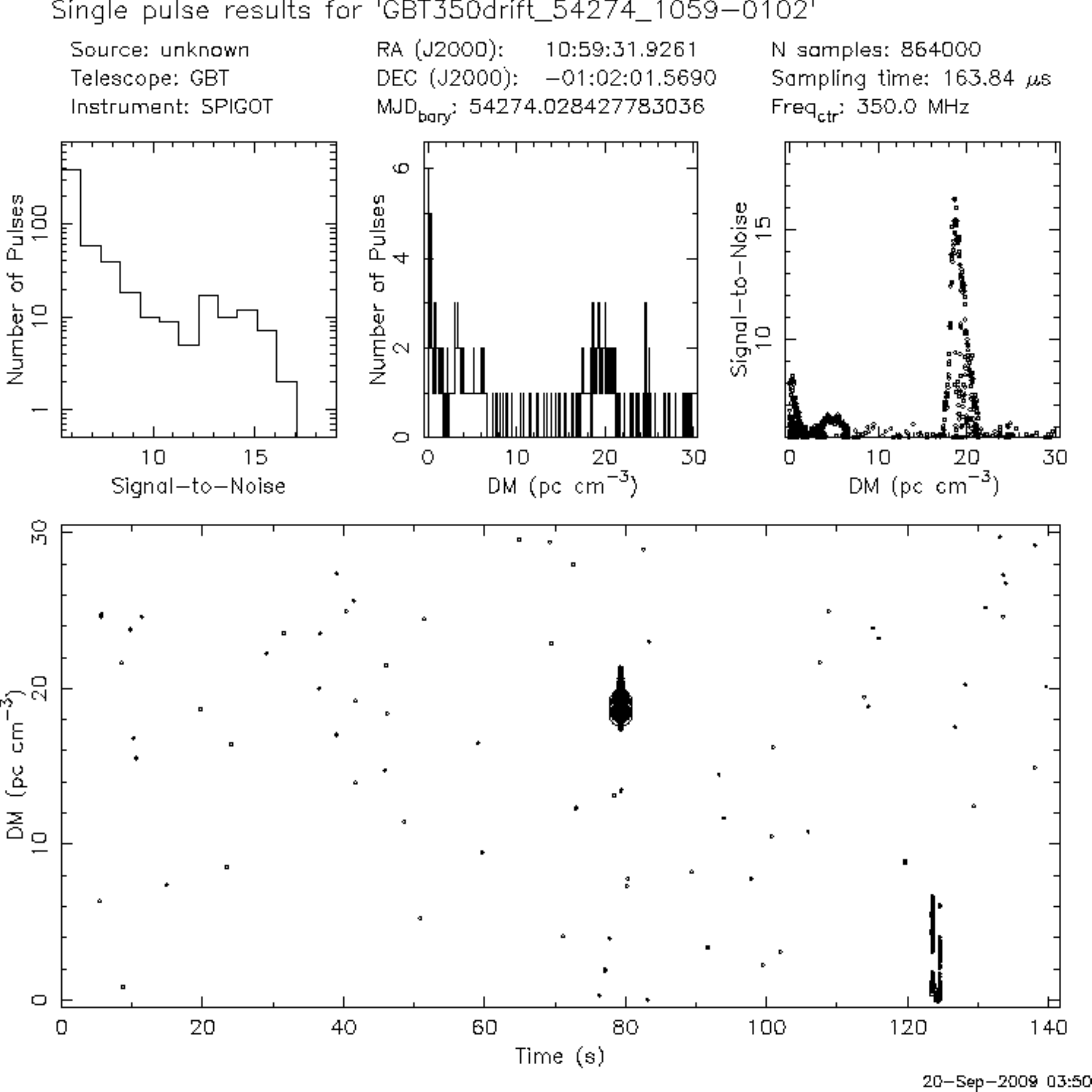}
    \\
    \vspace{0.5cm}
    \includegraphics[width=0.4\linewidth]{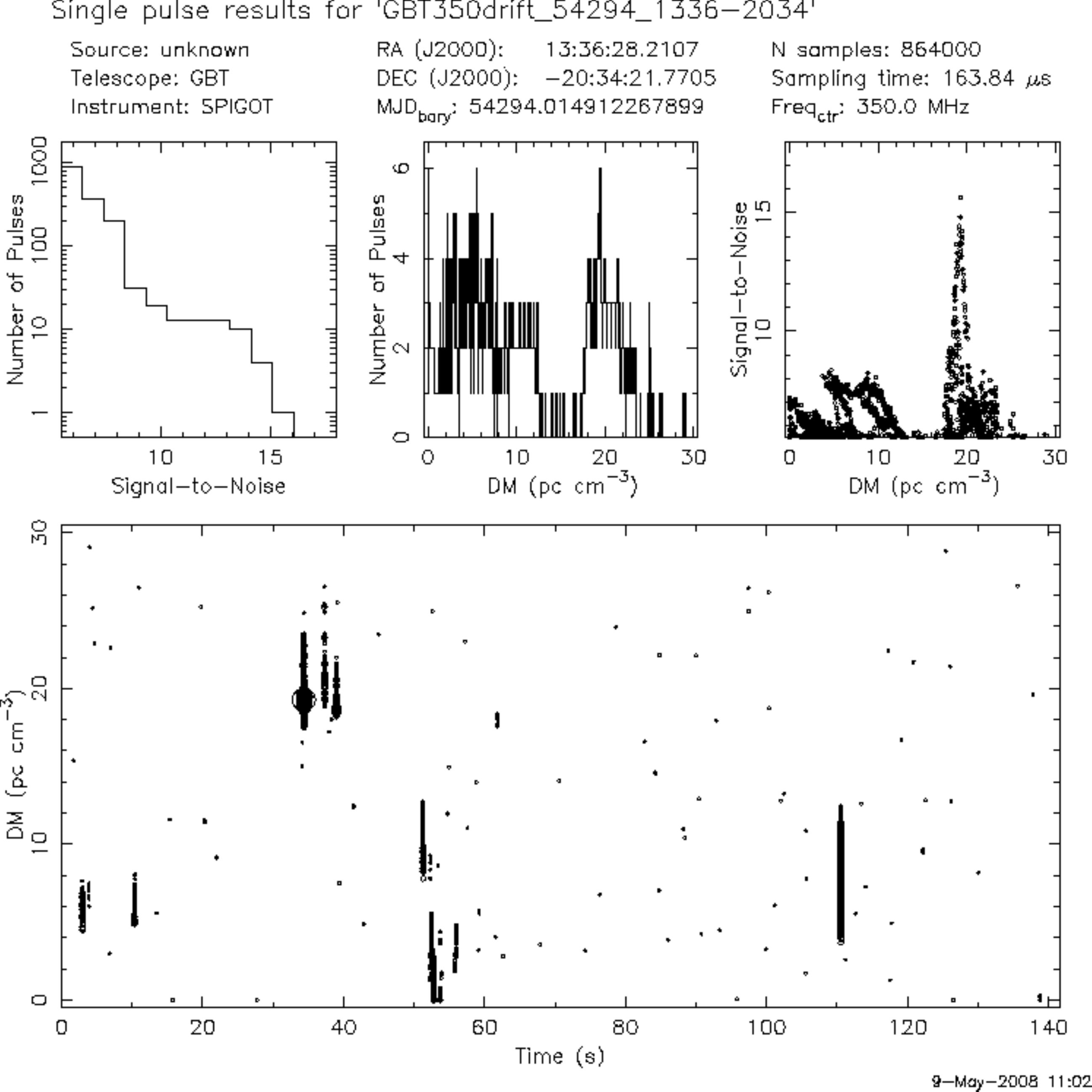}
    \includegraphics[width=0.4\linewidth]{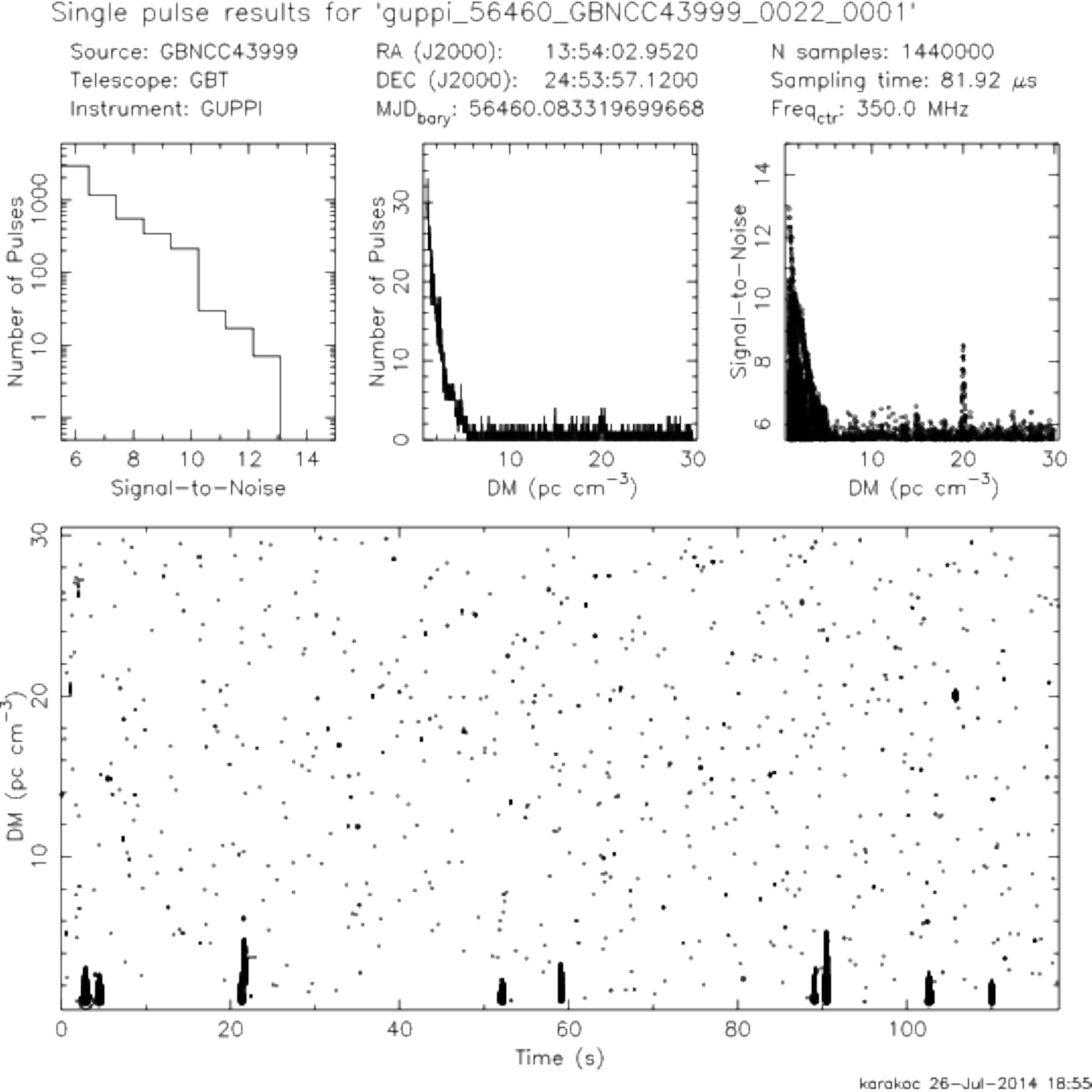}
    \caption{Discovery plots of the six unconfirmed RRAT candidates listed in Table~\ref{table:rratcands}. From left to right and top to bottom, these are candidates J0441$-$04, J0513$-$04, J0614$-$03, J1059$-$01, J1336$-$20, and J1354+25. Each plot is made up of four subplots, showing the number of events vs. S/N, the number of events vs. DM, the S/N vs. DM, and DM vs. time of detected events, with the plotted circle size corresponding to each event's S/N.\label{fig:rratcands}}
\end{figure*}

We applied our RRAT search algorithm to the Drift-scan survey, and to 21300 beams of the GBNCC survey, representing about a third of the GBNCC data taken thus far. This resulted in the discovery of 21 new RRATs and 6 RRAT candidates. 
The positions of these RRATs are indicated on the survey sky coverage map, in Figure~\ref{fig:skycoverage}. We present the properties of each RRAT in Table~\ref{table:rratdiscoveries}, including position, period, DM, and the survey in which it was found. 
We note that these parameters were obtained through follow-up observations as will be described in Section~\ref{sec:followup}, and the refined values are the ones reported. 
The RRAT candidates, as well as the number and duration of re-observations for each one, are listed in Table~\ref{table:rratcands}. The discovery plots for the candidates are shown in Figure~\ref{fig:rratcands}. We note that these candidates are all fairly close in DM; some of them may therefore be caused by RFI, or may simply be coincidentally close in DM.  
We define a candidate as any source that was only detected in the initial survey observation, and was not re-detected in any observations thereafter. These could thus be spurious astrophysical signals, or they could be genuine RRATs with low burst rates. 

\begin{deluxetable*}{lccccc} % if 2 columns
    \tablecaption{New RRAT Discoveries in Green Bank Telescope 350~MHz Surveys.\label{table:rratdiscoveries}}
    \tablewidth{0pt}
    \tablehead{\colhead{RRAT} & \colhead{R.A.} & \colhead{Dec.} & \colhead{P} & \colhead{DM} & \colhead{Survey}\\
                              & \colhead{(hh:mm:ss)}    & \colhead{(dd:mm)}&\colhead{(s)}&\colhead{(\dmunit)}& }
    \startdata
      J0054+69     & 00:54:28(25) & +69:26(2)   & ...  & 90.3(2)       & GBNCC \\ %pos:LOFAR heatmap (uncertainty too)
      J0103+54     & 01:03:37(20)  & +54:02(5)  & 0.354304(6) & 55.605(4) & GBNCC\\ %pos:LOFAR
      J0201+7005$^{*}$ & 02:01:41.344(7) & +70:05:18.11(6) & 1.349184471847(9)& 21.029(2) & GBNCC\\ %pos:timing
      J0332+79     & 03:32:45(20) & +79:10(5) & 2.05621(6) & 16.67(2) & GBNCC \\ %pos:LOFAR
      J0447$-$04   & 04:47(1) & $-$04:35(18) & 2.18819(2) & 29.83(4) & Drift \\ %pos:GBT820
      J0545$-$03   & 05:45(2) & $-$03:10(36) & 1.07393(2) & 67.2(4)  & Drift \\ %pos:GBT350
      J0957$-$06   & 09:57(2) & $-$06:17(36) & 1.72370(8) & 26.95(2) & Drift \\ %pos:GBT350
      J1126$-$27   & 11:26(2) & $-$27:37(36) & 0.358161(8)& 26.860(7)& GBNCC \\ %pos:GBT350
      J1153$-$21   & 11:53(2) & $-$21:18(36) & 2.34348(5) & 34.8(1)  & GBNCC \\ %pos:GBT350
      J1332$-$03   & 13:32(2) & $-$03:26(36) & 1.10640(5)  & 27.1(2) & Drift \\ %pos:GBT350
      J1439+76     & 14:39(2)  & +76:55(36)  & 0.947903(2) & 22.29(2)& GBNCC \\ %pos:GBT350
      J1538+2345$^{*}$ & 15:38:06.07(2) & +23:45:04.0(2)  &3.44938495332(9)& 14.909(1)& Drift \\ %pos:timing
      J1611$-$01   & 16:11(2) & $-$01:28(36) & 1.29687(2) & 27.21(7) & Drift \\ %pos:GBT350
      J1705$-$04   & 17:05(2) & $-$04:41(36) & 0.23748(2) & 42.951(9) & Drift \\ %pos:GBT350
      J1915$-$11   & 19:15(2) & $-$11:30(36) & 2.1770(2) & 91.06(8)  & Drift \\ %pos:GBT350
      J1944$-$10   & 19:44(2) & $-$10:17(36) & 0.409135(1)& 31.01(3)  & Drift \\ %pos:GBT350
      J1956$-$28   & 19:56(2) & $-$27:53(36) & 0.2600144(6)& 45.69(1) & GBNCC \\ %pos:GBT350
      J2007+20     & 20:07(2) & +20:21(36)   & 4.634(7) & 67.0(4)     & Drift \\ %pos:GBT350
      J2105+6223$^{*}$ & 21:05:12.93(2)  & +62:23:05.5(1) & 2.30487883766(4) & 50.75(8)    & GBNCC \\ %pos:timing
      J2311+67     & 23:11(1) & +67:05(18) & 1.9447(2) & 97.1(2) & GBNCC \\ %pos:GBT820
      J2325$-$0530$^{*}$ & 23:25:15.3(1) & $-$05:30:39(4) & 0.868735115026(9)& 14.966(7) & Drift \\
    \enddata
    \tablecomments{Periods that could not be reliably constrained are not reported (see text). Uncertainties on measurements are given in brackets, representing the uncertainty on the last reported digit. Some source positions are better determined than others thanks to position refinement observations, as described in Section~\ref{subsec:gridding}, and others are even better determined using timing (see Section~\ref{subsec:timedrrats}). Sources for which we have full timing solutions are indicated by an asterisk and are presented in Table~\ref{table:timingparams}.}
%\end{deluxetable}
\end{deluxetable*}
%position sigfigs and uncertainty: GBT350: 0.6deg=36'=2.4m, GBT820: 0.3deg=18'=1.2m, LOFAR: 5'=20s, so RA: hh:mm for GB, hh:mm:ss for LOFAR, and dec: dd:mm for all three
% converting uncertainties to RA and dec:
% LOFAR: 5' = 20s
% 1 hr = 15 deg, so (5'*1deg/60')*1hr/15deg*3600s/hr = 20 s

\begin{deluxetable*}{lcccccccc} % if 2 columns
    \tablecaption{Unconfirmed RRAT Candidates in Green Bank Telescope 350~MHz Surveys.\label{table:rratcands}}
    \tablewidth{0pt}
    \tablehead{\colhead{RRAT} & \colhead{R.A.} & \colhead{Dec.} & \colhead{P} & \colhead{DM} & \colhead{Number of pulses} & \colhead{Peak flux density} & \colhead{Survey} & Re-observations\\
                              & \colhead{(hh:mm)} & \colhead{(dd:mm)}& \colhead{(s)} &\colhead{(\dmunit)} & & (Jy) & &(min)}
    \startdata
      J0441$-$04 & 04:41 & $-$04:18 &...   & 20.0 & 1 & 0.4 & Drift & $15+3+15+10$\\
      J0513$-$04 & 05:13 & $-$04:18 &...   & 18.5 & 1 & 1.0 & Drift & $10+10+10$ \\
      J0614$-$03 & 06:15 & $-$03:29 & 0.136& 17.9 & 3 & 0.9 & Drift& $6+10+10$ \\
      J1059$-$01 & 10:59 & $-$01:02 &...   & 18.7 & 1 & 0.8 & Drift& $10+10.5$ \\
      J1336$-$20 & 13:36 & $-$20:34 & 0.184& 19.3 & 3 & 0.8 & Drift& $10+10+10$ \\
      J1354+25   & 13:54 &   +24:54 & ...  & 20.0 & 1 & 0.1 & GBNCC& 10 \\
    \enddata
    \tablecomments{True period values may be integer fractions of the tabulated values, as described in Section~\ref{subsec:timing}. Periods for sources that showed two or fewer pulses per observation could not be reliably constrained and are not reported. No detailed period and DM refinement, nor a robust determination of uncertainties, was done for these sources since they are pending confirmation. The peak flux density of each candidate was computed using the observed pulse width (assumed to be the optimal box-car width used in the search), maximum S/N, and telescope parameters. A summation of integration times in the `Re-observations' column indicates separate observation epochs.}
\end{deluxetable*}

\section{RRAT follow-up observations and analysis}
\label{sec:followup}
We conducted follow-up observations of our RRAT discoveries between February 2012 and January 2015 using the GBT and LOFAR telescopes. Observations with the GBT were conducted at center frequencies of 350~MHz and 820~MHz, with bandwidths of 100~MHz and 200~MHz, respectively. Integration times ranged from 6--15 minutes. Observations with LOFAR used a center frequency of 150~MHz, a bandwidth of 80~MHz, and integration times of 10--30 minutes. These and other observation parameters are summarized in Table~\ref{table:obsmodes} for all observing modes used. These observations allowed us to refine positions, periods, DMs, and burst rates for our RRATs, using the methods described below. We also obtained phase-coherent timing solutions for four RRATs. We describe this timing analysis below and present the results in Table~\ref{table:timingparams}. 

LOFAR is made up of many dipole antennas which are arranged hierarchically in stations, with core stations densely packed near Exloo, the Netherlands, remote stations spread throughout the Netherlands, and international stations in Germany, France, Sweden, and the United Kingdom. For our LOFAR observations, we used all core high-band antenna (HBA; 110--190 MHz) stations. 
The LOFAR core stations subtend a maximum baseline of approximately 2~km and create a tied-array beam with a FWHM of approximately 5$'$ (at 150~MHz). Typically 20-23 core stations were used in each LOFAR observation.
When observing using multiple stations, one can take advantage of multi-beaming in order to form and observe several beams on the sky simultaneously. Moreover, adding signals from stations coherently results in a tied-array beam, which has high spatial resolution and increased sensitivity. We used these configurations for our observations with LOFAR, as we describe below. More details on available observing modes with LOFAR are presented in \cite{hwg+13} and \cite{sha+11}.

\subsection{Position refinement}
\label{subsec:gridding}
\begin{figure}
\centering
\includegraphics[width=\linewidth]{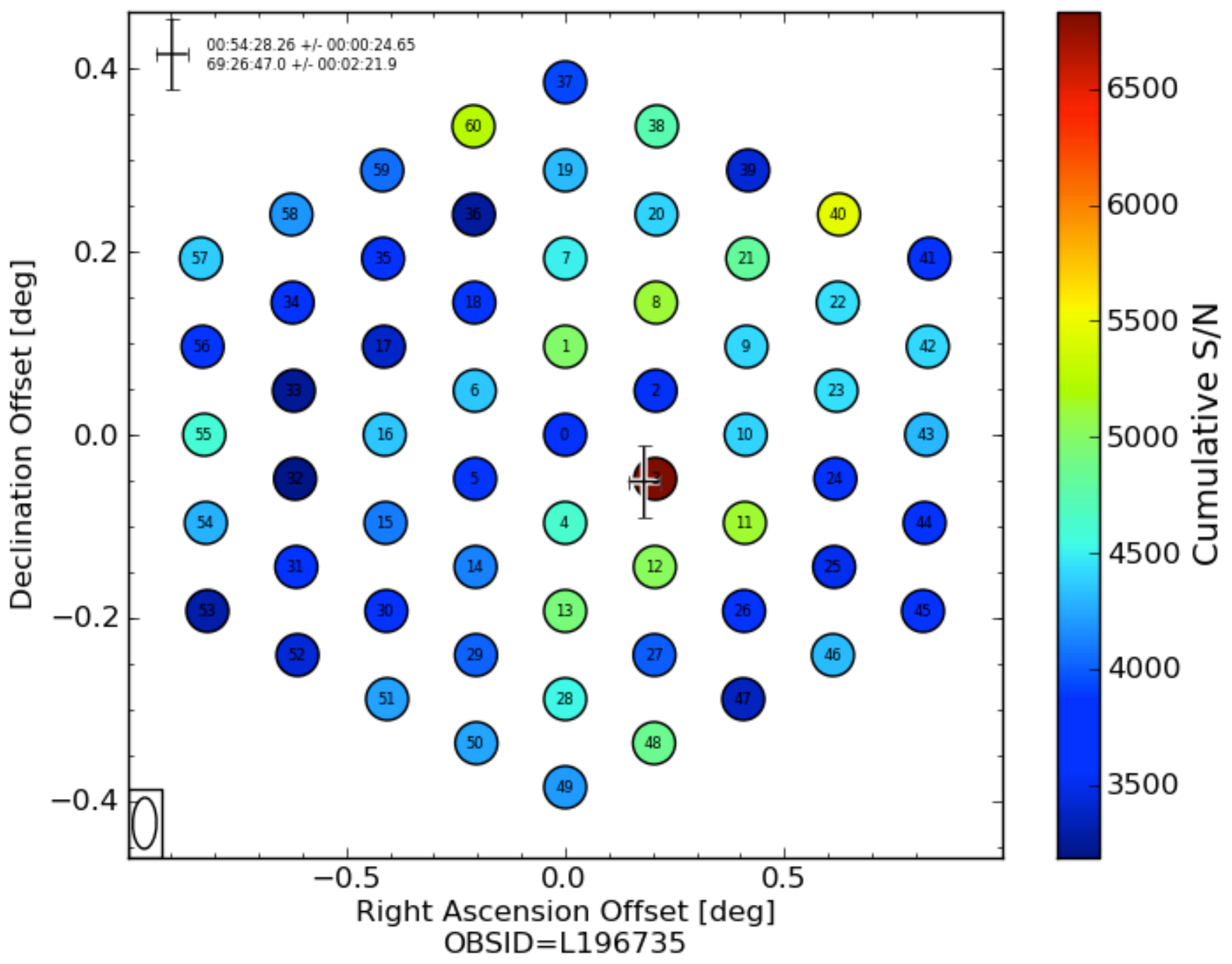}
\caption{Single-pulse detection heatmap for LOFAR gridding observation of RRAT J0054+69, taken December 20, 2013, as part of our follow-up campaign. The heatmap compares detection significance of single-pulse events at the RRAT's DM across all beams. The S/N values are summed for all events at the given DM, for each beam. A detection is evident in beam 3. The apparent detections in some of the other beams (e.g. 11, 40, 60) are due to RFI signals. Note that the circle sizes do not represent the size of the tied-array beams on the sky. The synthesized beam is shown in the bottom-left corner.\label{fig:0053_lofar_heatmap_sp}}
\end{figure}

Positions were refined using standard gridding with the GBT at 820~MHz for some sources, and using LOFAR beamforming for other sources, reducing positional uncertainties to 15$'$ (GBT) or 5$'$ (LOFAR). In particular, LOFAR was especially useful for the gridding observations, since we were able to use all core high-band antenna stations to form 61 simultaneous tied-array beams, each of width 5$'$, and arranged in four concentric rings, thus efficiently covering the 350-MHz GBT beam. 
This setup is particularly useful for localizing RRATs, compared to sequential gridding observations which require a longer integration time overall and are made challenging by RRATs' sporadic emission and pulse amplitude variations. 
Figure~\ref{fig:0053_lofar_heatmap_sp} shows an example of the LOFAR configuration for gridding observations of one of the RRATs.

\begin{figure}
\centering
\includegraphics[width=\linewidth]{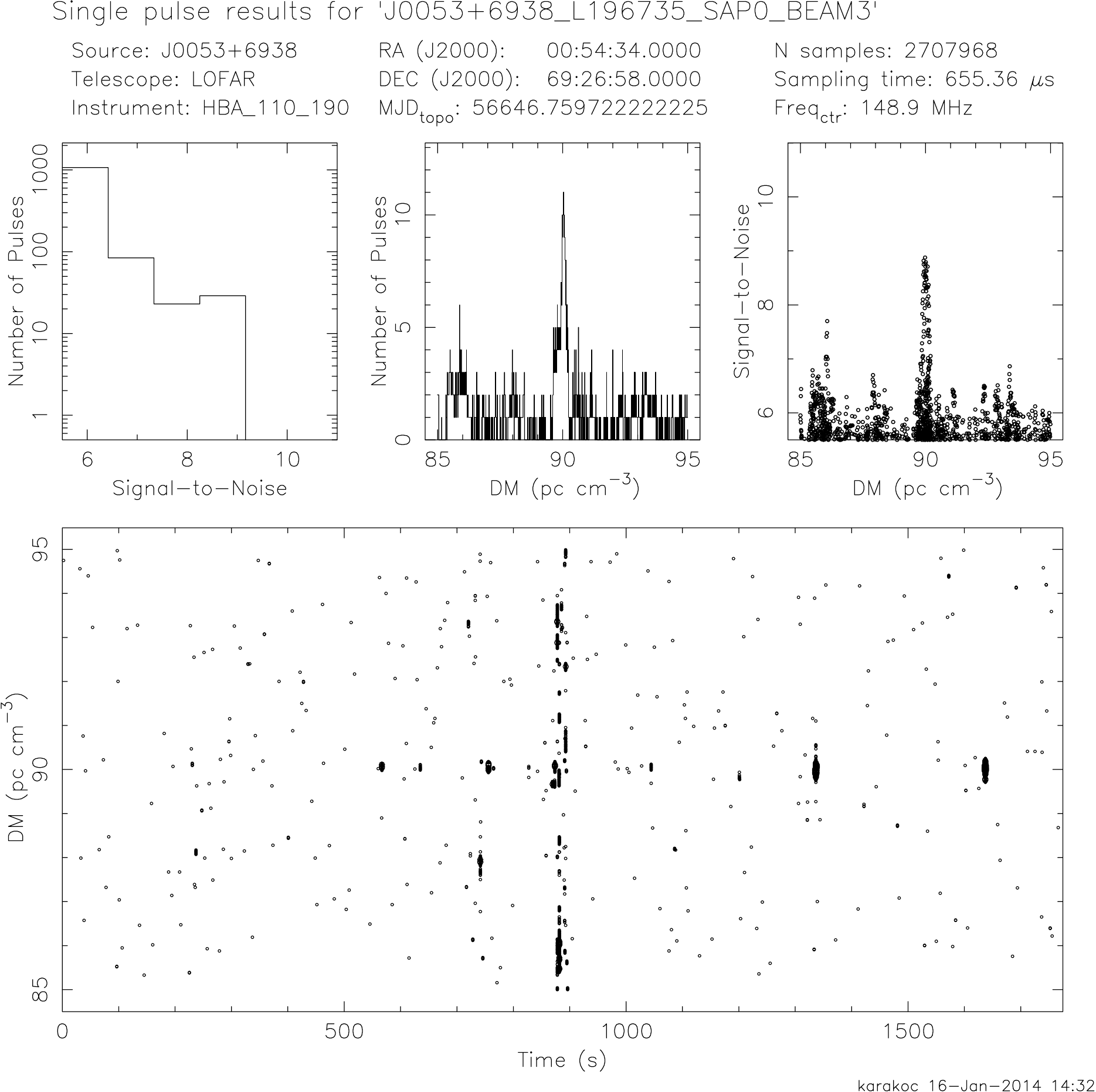}
\caption{LOFAR detection of single pulses from RRAT J0054+69. Approximately 7 pulses at a DM of 90~$\mathrm{pc\, cm}^{-3}$ are detected during a gridding observation, allowing the refinement of the position of this source to that of beam 3 of the 61 tied-array beam grid, thus lowering the position uncertainty from 36$'$ to 5$'$. A bright RFI signal is seen at $t\sim900$~s.\label{fig:0053_lofar_sp}}
\end{figure}

In the analysis of LOFAR gridding observations, each beam is processed independently using standard pulsar analysis tools, and summary plots are then produced for the observation to present results for all tied-array beams. These diagnostic plots include detection `heatmaps' of two types: The periodicity heatmap shows the relative detection significance when each beam is folded at the nominal period of the RRAT, and the RRAT heatmap compares the single-pulse detection strength at the DM of the RRAT in each beam. An example of the single-pulse heatmap for the observation of RRAT J0054+69 is seen in Figure~\ref{fig:0053_lofar_heatmap_sp}, and the corresponding detection is shown in Figure~\ref{fig:0053_lofar_sp}. 
The RRAT heatmap provides an overview of the single-pulse results without requiring the inspection of a single-pulse plot for every beam, though the latter may still be needed when the detection is weak or when RFI is present.

\subsection{Burst rate calculations}
\label{subsec:br}
An important characteristic of RRATs, and one of the properties that determines whether or not a source can be timed, is the source burst rate. For every observation of a given source, we compute the burst rate by dividing the number of detected pulses by the observation duration.
Note that observed burst rates are observing-setup-dependent, as a telescope's sensitivity will affect its ability to detect weaker pulses, and single pulses from a given source vary in amplitude \citep[e.g.][]{bjs+12}. RRATs may also behave differently at different frequencies, depending on their spectrum. It is generally expected that RRATs, like pulsars, are broadband sources, and emit over the entire radio spectrum. However, if one finds significantly different burst rates at different frequencies, some inferences regarding the source spectrum can be made.

Moreover, since RRATs are erratic emitters, their measured burst rates may change drastically between observations using the same setup. It is thus important to use data from multiple, sufficiently long observations, and report the average burst rate, computed by dividing the number of detected pulses over all observations by the summed observation time. We estimate the uncertainty on the number of detected pulses using Poisson statistics. 

A summary of source detections and burst rates with each observing setup is presented in Table~\ref{table:burstrates}. 
Since LOFAR's sensitivity is elevation-dependent, the LOFAR burst rates of different RRATs are not directly comparable to one another.
Moreover, many RRATs were not detected with LOFAR, which may be due to reduced sensitivity at low elevations (those sources with low declinations only reach a maximum elevation of $\sim$$30^{\circ}$ at LOFAR), higher sky temperatures close to the Galactic Plane, or their spectra.

\subsection{Timing and period refinement}
\label{subsec:timing}
When RRATs are initially discovered, their periods can be constrained by finding the greatest common divisor of the time differences between detected pulses. This can typically be done as long as three or more pulses are observed. The obtained value may then be an integer multiple of the true period, but as more pulses are observed we can establish a period value with higher confidence.  
These period values are then refined through follow-up observations and timing techniques, as described below. Furthermore, if a sufficiently long data set is available, we can derive a phase-coherent timing solution. 

We have timed four RRATs, and refined the single-pulse-inferred period values for the other sources. 
For one RRAT, J0054+69, we observed an abundance of pulses, however we were unable to fit a periodicity to these pulses. We suspect that this may be due to a complex pulse profile, with detected single pulses corresponding to different profile components. In fact, even for single-peaked average profiles, individual pulses from a RRAT are narrower and distributed across the envelope of the average pulse profile, introducing jitter which makes timing difficult.

We start by extracting all single pulses from the data. 
The output of \spsearch provides approximate times of detected pulses. 
One way of proceeding, then, for a source whose period is roughly known, is to divide the observation into time segments corresponding to single rotations (or `pulse archives'), and examine those data segments near the approximate pulse times by eye until identifying bright pulses. We perform the data chopping using the {\tt DSPSR} tool \citep{sb11}, which also allows to simultaneously dedisperse the data according to the RRAT's DM. We then use {\tt PSRCHIVE} tools \citep{hsm04} to make frequency-versus-phase plots for each pulse archive, where pulses can be identified.

Once all single pulses for a given observation have been identified, we align the pulses by cross-correlation and add them together using {\tt psradd} to create a high S/N pulse. We then use {\tt paas} to interactively fit an analytic pulse profile template to the combined pulse. We use this as an initial template for the remaining observations (if available), but further refine the template once a timing solution is obtained (see Section~\ref{subsec:timedrrats}).
A time-of-arrival (TOA) is then computed for each single pulse, using the standard cross-correlation algorithm in the {\tt pat} tool \citep{tay92}.
After extracting TOAs for each source, we use \tempotwo\footnote{\url{http://www.atnf.csiro.au/research/pulsar/tempo2}} and standard pulsar timing techniques to refine the period within a single observation for each source. 

We used these techniques to refine the periods of the 14 RRATs for which we did not have many observations. We used a similar approach but different software for the four RRATs for which we had a timing baseline of observations, as we describe in Section~\ref{subsec:timedrrats}.

\subsection{RRATs with timing solutions}
\label{subsec:timedrrats}
We obtained phase-coherent timing solutions for four RRATs using similar techniques to those described above.
For these RRATs, we had more data and pulse detections, and were thus able to obtain stable pulse profiles using high S/N observations or multiple observations summed together. We therefore did not need the fine-grain control offered by {\tt DSPSR} and {\tt PSRCHIVE}, and instead used \presto tools for this analysis. 
We folded the data with \prepfold while specifying a large number of subintegrations, each corresponding to a few seconds, generated TOAs using {\tt get\_TOAs.py}, and ignored all subintegrations that did not contain pulses.
We found that two of the four RRATs were consistently detectable from folding the data, but this method worked well for all four nonetheless. 
We then used \tempotwo for the timing. 

\begin{figure*} % for 2 columns
\centering
\includegraphics[width=0.8\linewidth]{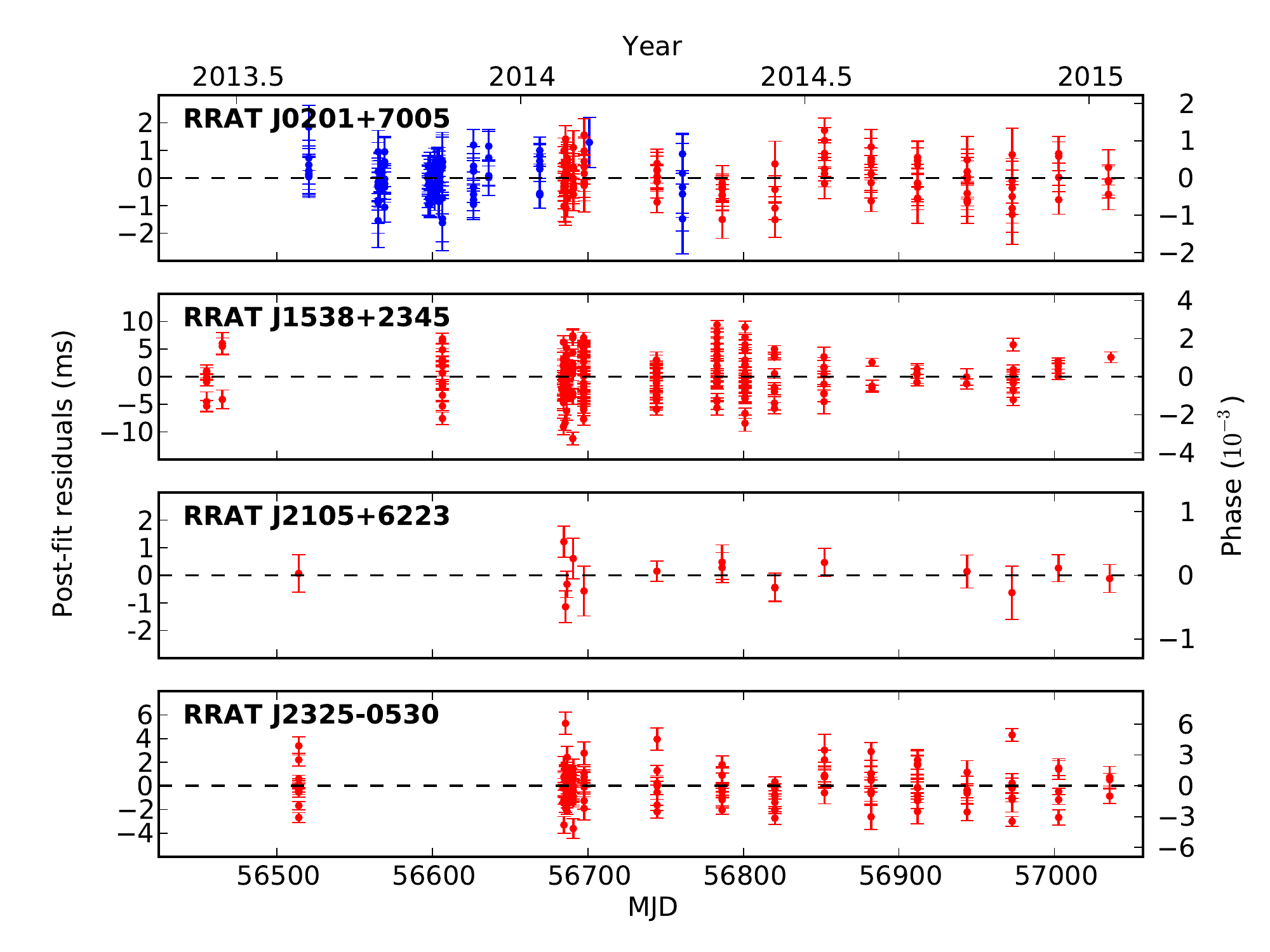}
\caption[]{Timing residuals for the four RRATs. Residuals from GBT 820-MHz data are plotted in blue, whereas those from LOFAR observations are plotted in red. Note the different scale for each subplot.\label{fig:residuals}}
\end{figure*}

\begin{figure*}
\centering
\includegraphics[width=0.75\linewidth]{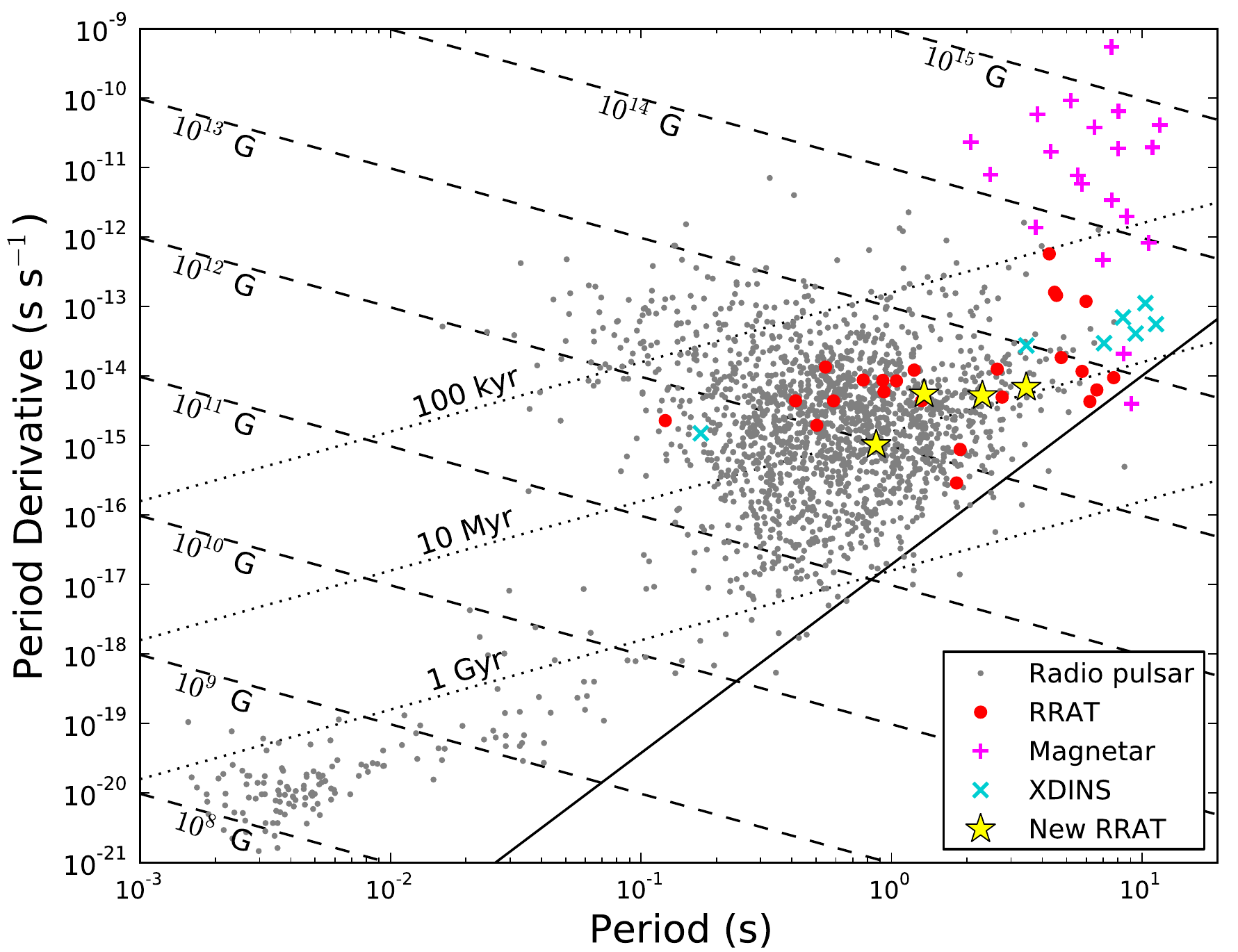}
\caption{$P-\dot{P}$ diagram for all known neutron stars outside of globular clusters. The four new RRATs are indicated by stars, the previously known RRATs with measured $\dot{P}$ by large circles, magnetars by crosses, XDINS by X's, and other neutron stars by dots. Lines of constant magnetic field are dashed and lines of constant characteristic age are dotted. The solid line represents the `death line' (using Equation~9 of \citealt{cr93}), beyond which we do not expect to find pulsars, but note that the exact location of the death line is uncertain since the emission mechanism of pulsars is not fully understood. Values for pulsars were taken from the ATNF Pulsar Catalogue \citep{mht+05}, for magnetars from the McGill Magnetar Catalog \citep{ok14}, and for RRATs from the RRATalog\footnote{\url{http://astro.phys.wvu.edu/rratalog}}. \label{fig:p-pdot}}
\end{figure*}

The timing solutions for these RRATs are presented in Table~\ref{table:timingparams}, and their residuals are shown in Figure~\ref{fig:residuals}. We also plot these RRATs on the $P-\dot{P}$ diagram in Figure~\ref{fig:p-pdot}. Comparing these four RRATs to the previously known RRATs, we see that they lie roughly in the same region, with periods of order 1~s and magnetic fields of order~$10^{12}$~G. 

We did not have sufficient data in order to obtain timing solutions for the other RRATs, and in particular were not able to detect many of them with LOFAR (as seen in Table~\ref{table:burstrates}). Given sufficient telescope time, we should be able to obtain timing solutions for the remaining RRATs, for example using the GBT. It is also possible that some of these sources would be detected in longer observations with LOFAR. There are also upcoming telescopes, such as the Canadian Hydrogen Intensity Mapping Experiment \citep[CHIME;][]{baa+14}, which will provide daily coverage of the sky and will thus be very useful for observing and timing RRATs.

\subsection{DM determination}
\label{subsec:dm}
The DM of a RRAT can be approximated by examining the S/N behavior of a group of single-pulse events associated with an individual pulse. As described in Section~\ref{subsubsec:algorithm design}, the S/N of a pulse will peak when the data are dedispersed at the optimal DM. We can thus estimate the DM as the value at which the RRAT's pulses are brightest. This method requires no additional computation as it uses data products produced by the data reduction pipeline. However, because the single-pulse search in the pipeline uses computationally efficient box-car filters as an approximation to the actual spiky shape of astrophysical pulses, the DM values determined from this method have some slight systematic errors that are difficult to quantify. 

In order to refine the DM of a RRAT, we first identify all single pulses using {\tt DSPSR} and {\tt PSRCHIVE} as described in Section~\ref{subsec:timing}, but without dedispersing the data. We then choose the brightest pulse, or a sum of several pulses, divide the frequency band into 4--6 subbands, and extract a TOA for each subband.
Due to the dispersion sweep, we expect that these TOAs will obey the $1/f^2$ plasma dispersion law. We then input these TOAs into \tempotwo and fit for the DM, yielding a refined DM value. 
We fit this DM value using one observation for each source. 

\begin{deluxetable*}{lcccc} 
    \tabletypesize{\footnotesize}
    \tablecaption{Timing parameters for four RRATs.\label{table:timingparams}}
    \tablewidth{0pt}
    \tablehead{ & \colhead{RRAT J0201+7005} & \colhead{RRAT J1538+2345} & \colhead{RRAT J2105+6223} & \colhead{RRAT J2325$-$0530}}
    \startdata
    \multicolumn{5}{c}{Measured parameters} \\
    \hline
    Right ascension (J2000)                   & 02:01:41.344(7)         & 15:38:06.07(2) & 21:05:12.93(2) & 23:25:15.3(1) \\
    Declination (J2000)                       & +70:05:18.11(6)         & +23:45:04.0(2) & +62:23:05.5(1) & $-$05:30:39(4) \\
    Spin frequency (Hz)                       & 0.741188488948(5)       &0.289906755417(8)& 0.433862285366(7) & 1.15109885937(1)\\
    Spin frequency derivative (Hz/s)          & $-3.0294(8)\times10^{-15}$& $-5.79(1)\times10^{-16}$ & $-9.824(1)\times10^{-16}$ &$-1.363(2)\times10^{-15}$\\
    Dispersion measure ($\mathrm{pc\, cm}^{-3}$) & 21.029(2)               & 14.909(1)      & 50.75(8)       & 14.966(7)   \\
    \hline
    \multicolumn{5}{c}{Derived parameters} \\
    \hline
    Galactic longitude, $l$ ($^{\circ}$)        & 128.89                  & 37.32         & 99.79 & 75.58    \\
    Galactic latitude, $b$ ($^{\circ}$)         & 8.03                    & 52.39         & 10.18 & $-60.20$ \\
    Spin period, $P$ (s)                        & 1.349184471847(9)       &3.44938495332(9)& 2.30487883766(4) & 0.868735115026(9)\\
    Spin-down rate, $\dot{P}$ (s/s)             & $5.514(1)\times10^{-15}$ & $6.89(1)\times10^{-15}$ & $5.219(6)\times10^{-15}$ & $1.029(2)\times10^{-15}$ \\
    Magnetic field, $B$ (G)                      & $2.8\times10^{12}$ & $4.9\times10^{12}$ & $3.5\times10^{12}$ & $9.6\times10^{11}$ \\
    Spin-down luminosity, $\dot{E}$ (erg s$^{-1}$)& $8.9\times10^{31}$ & $6.6\times10^{30}$  & $1.7\times10^{31}$ & $6.2\times10^{31}$ \\
    Characteristic age, $\tau_\mathrm{c}$ (yr)   & $3.9\times10^{6}$   & $7.9\times10^{6}$   & $7.0\times10^{6}$    & $1.3\times10^{7}$\\
    DM distance, $D$ (kpc)$^{\mathrm{a}}$        & 1.1                 &  1.2        & 2.6  &  0.7     \\
    \hline
    \multicolumn{5}{c}{Observation parameters} \\
    \hline
    Discovery (MJD)                              & 55163               & 54243    & 55201  & 54240\\
    Timing epoch (MJD)                           & 56777               & 56745    & 56774  & 56774\\
    Start epoch (MJD)                            & 56521               & 56455    & 56514  & 56514\\
    Finish epoch (MJD)                           & 57035               & 57036    & 57036  & 57036\\
    Number of TOAs                               & 204                 & 213      & 16     & 132     \\
    RMS post-fit residuals ($\mu$s)              & 610                 & 3203     & 523    & 1165
    \enddata
    \tablecomments{$^{\mathrm{a}}$ DM distances are as implied by the \cite{cl02} {\tt NE2001} model and have typical uncertainties of 25\%.}
\end{deluxetable*}
% Table up-to-date with TOAs up to (and incl) Jan obs for 0203, 1537, 2106, 2324.

% BURST RATES AND INTEGRATIONS TIMES - MERGED TABLE
\begin{deluxetable*}{lcccccccc}
    \tablecaption{RRAT burst rates from discovery and follow-up observations with the GBT and LOFAR telescopes.\label{table:burstrates}}
    \tablewidth{0pt}
    \tablehead{\colhead{RRAT} & \multicolumn{2}{c}{Discovery} & \multicolumn{2}{c}{GBT 350} & \multicolumn{2}{c}{GBT 820} & \multicolumn{2}{c}{LOFAR} \\
    & \colhead{Burst rate (hr$^{-1}$)} & \colhead{$t$ (min)} & \colhead{Burst rate (hr$^{-1}$)} & \colhead{$t$ (min)} & \colhead{Burst rate (hr$^{-1}$)} & \colhead{$t$ (min)} & \colhead{Burst rate (hr$^{-1}$)} & \colhead{$t$ (min)}}
    \startdata
    J0054+69    &300 $\pm$ 95& 2 & 186 $\pm$ 33& 10 &285 $\pm$ 21&40 & 6 $\pm$ 2&70\\ 
    J0103+54    &390 $\pm$ 108&2 & 60 $\pm$ 15& 15 &  0        & 30 & 3 $\pm$ 2&45 \\ 
    J0201+7005  &180 $\pm$ 73& 2 & 24 $\pm$ 10& 15 &256 $\pm$ 22& 30 & 63 $\pm$ 7&75\\ 
    J0332+79    & 30 $\pm$ 30& 2 &65 $\pm$ 12 & 25 &66 $\pm$ 10& 40 & 14 $\pm$ 3&70\\ 
    J0447$-$04  &103 $\pm$ 51& 2 &41 $\pm$ 10  &25 &78 $\pm$ 22& 10 &  0       &15\\ 
    J0545$-$03  & 77 $\pm$ 45 & 2 & 42 $\pm$ 16 &10 &  --       & -- &   0       &15 \\ 
    J0957$-$06  &180 $\pm$ 68& 2 & 138 $\pm$ 29&10 &   --      & -- &  0        &10\\ 
    J1126$-$27  &180 $\pm$ 73 & 2 & 48 $\pm$ 17 &10 & --        & -- & --        &--\\
    J1153$-$21  &150 $\pm$67 & 2 & 60 $\pm$ 24 &10 &   --      & -- &  --       &--\\ 
    J1332$-$03  & 51 $\pm$ 36& 2 & 20 $\pm$ 5  &40 &  --       & -- &  --       &--\\ 
    J1439+76    &450 $\pm$ 116&2 & 406 $\pm$ 48& 11&   0       & 7  &  0       &15\\ 
    J1538+2345  &129 $\pm$ 58&2  & 77 $\pm$ 14 & 23 &   --      & -- &66 $\pm$ 7&75 \\     
    J1611$-$01  & 51 $\pm$ 36& 2 & 60 $\pm$ 19 &10 &   --      & -- &  0        &15\\ 
    J1705$-$04  & 26 $\pm$ 26& 2 & 32 $\pm$ 11 &15 &   --      & -- &  0        &15\\ 
    J1915$-$11  & 26 $\pm$ 26& 2 &108 $\pm$ 25& 10 &  --       & -- &  0        &15 \\ 
    J1944$-$10  &180 $\pm$ 68& 2 & 78 $\pm$ 13 &27 &   --      & -- &  0        &15 \\ 
    J1956$-$28  &120 $\pm$ 60& 2 & 66 $\pm$ 20 &10 &   --      & -- &  --       &--\\ 
    J2007+20    & 77 $\pm$ 44& 2 & 11 $\pm$ 4  & 34&   --      & -- &  0        &15 \\ 
    J2105+6223  & 30 $\pm$ 30& 2 &  --        & -- &  --       & --  & 12 $\pm$ 3 &75  \\ 
    J2311+67    & 60 $\pm$ 42& 2 &  --        & -- &168 $\pm$ 32& 10 &  0        &75 \\ 
    J2325$-$0530 &103 $\pm$ 51&2 & 46 $\pm$ 9 & 33 & --        & --  &52 $\pm$ 8&45  \\ 
    \enddata
    \tablecomments{A field containing `--' indicates the source was not observed for follow-up with that setup, whereas 0 indicates the source was observed and not detected. Average source burst rates and uncertainties are computed as described in the text. Note that the observed burst rate for a given source may vary significantly between different observing setups, as described in Section~\ref{subsec:br}.}  
\end{deluxetable*}

\section{Discussion and population comparisons}
\label{sec:discussion}
We now consider our new RRATs in the context of the previously known RRAT population, as well as the general radio pulsar population. We examine the observed burst rates for our RRATs, and discuss their implications on survey detections. We then compare properties such as period and DM for RRATs and pulsars, in order to assert whether the underlying distributions are statistically different. We consider the relevant detection selection biases, and describe how to conduct fair statistical comparisons despite them. 
Finally, we discuss the fact that we did not find any FRBs with relation to the published FRB rates.

\subsection{Burst rates}
\label{subsec:discussion br}
\begin{figure}
\centering
\includegraphics[width=\linewidth]{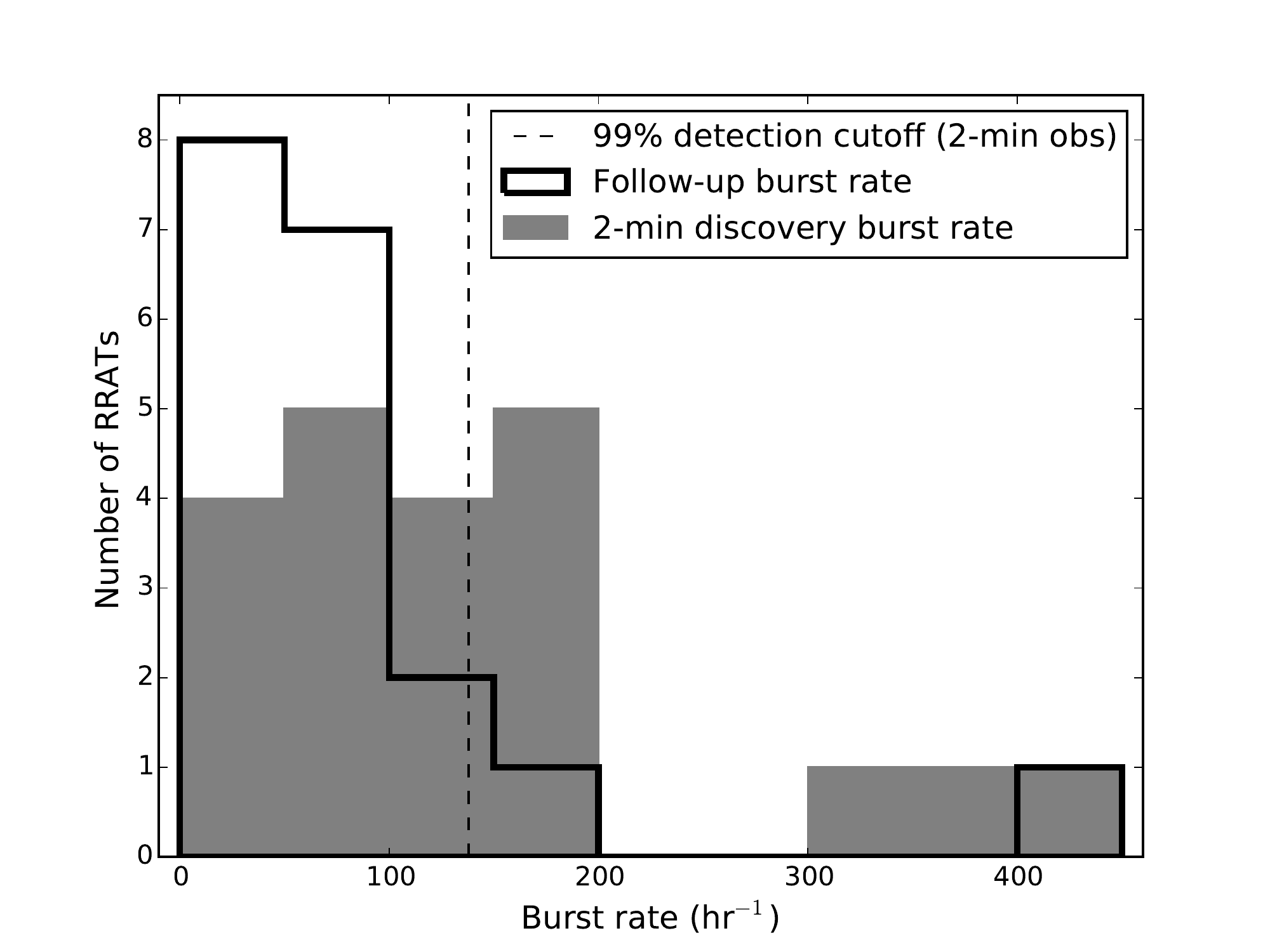}
\caption[Burst rate distribution of new RRATs]{Burst rate distribution of our new RRATs. The rates plotted are those inferred from the GBT~350-MHz discovery and follow-up observations, corresponding to columns~2 and~3 of Table~\ref{table:burstrates}, respectively. The 99\% Poisson burst rate cutoff is also shown, indicating the rate below which it is unlikely to detect RRATs in a 2-min observation.\label{fig:BRhist}}
\end{figure}

As is evident from the new RRATs' burst rates, which are recorded in Table~\ref{table:burstrates} and plotted in Figure~\ref{fig:BRhist}, there is a distribution of rates with values varying drastically from source to source.  
In fact, even for a given source, the measured burst rate may vary between observations due to small-number (Poisson) statistics, and the expected rate also varies with the observation's sensitivity (including factors such as observing setup, source elevation, and RFI occupancy). It is therefore difficult to isolate the intrinsic burst rate from the observed rate, and to know whether we have found the true average burst rate. The reported values should therefore be considered approximate. 

These modest burst rates demonstrate the potential of missing a RRAT in a short observation, and suggest that the number of RRATs detected in a given surveyed region is only a fraction of the number therein. 
By approximating RRAT pulse emission as a random process that follows Poisson statistics, we can estimate the true number of RRATs present within a certain region given the number of detected RRATs and an average burst rate for the population.

\begin{figure}
\centering
\includegraphics[width=\linewidth]{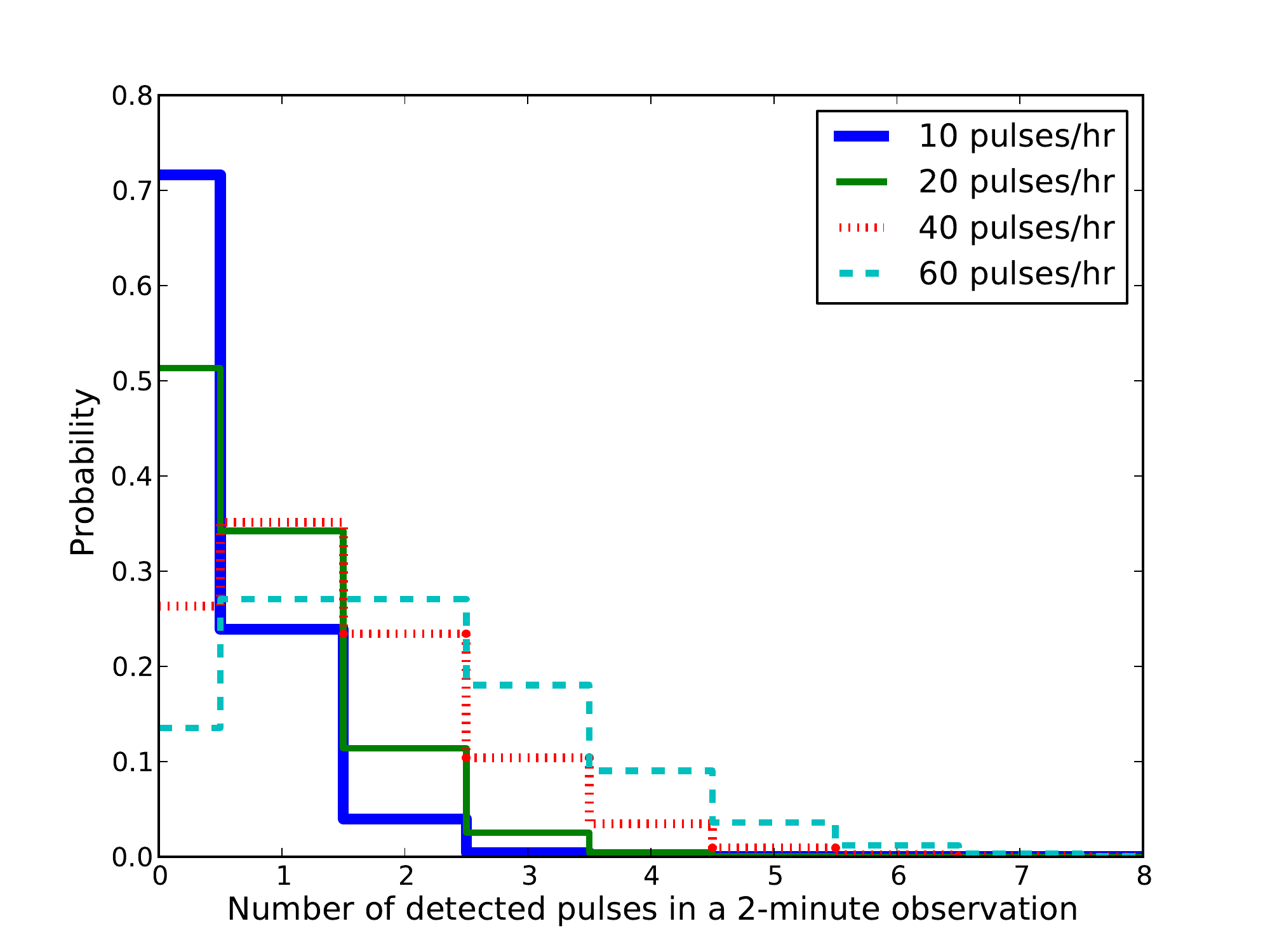}
\caption[Probability of RRAT pulse detection]{Poisson probability function showing the probability of pulse detection over a 2-min observation, for different assumed RRAT burst rates. \label{fig:poissonpdf}}
\end{figure}

The Poisson probability function is given by
\begin{equation}
\label{eq:poissonpdf}
\mathrm{Pr}(X = k) = \frac{\lambda^k e^{-\lambda}}{k!}\, ,
\end{equation}
where, in this case, $X$ is the number of detected pulses, $\mathrm{Pr}(X = k)$ is the probability of detecting $k$ pulses, and $\lambda$ is the expected value of $X$ given by the product of observation duration and mean burst rate. The probability functions for different mean burst rates are plotted in Figure~\ref{fig:poissonpdf}, where we have assumed a typical 2-min survey observation.
For each burst rate, the probability of missing a RRAT over an observation is given by $\mathrm{Pr}(X=0)$, whereas the probability of detecting one or more pulses is given by the sum $\sum\nolimits_{k=1}^\infty \mathrm{Pr}(X = k)$, or simply $1-\mathrm{Pr}(X=0)$.

We can thus assume some mean burst rate for the RRAT population, and use the above to compute the probability of non-detection. In this way, we obtain an estimate of the fraction of the RRAT population missed in a given pulsar survey due to short observations and infrequent emission. 
It should be noted however that the observed burst rate distribution for our sources is inherently biased, since these sources had sufficiently high burst rates to be detected. 
In fact, comparing the follow-up to discovery burst rate of each source, we see that the follow-up rates are generally lower, hinting that the survey burst rates are biased toward high values. 
We therefore cannot assume the average burst rate from our discoveries as that of the entire RRAT population. 
Instead, we only illustrate this using several RRAT burst rates.

With the observed burst rates tabulated in Table~\ref{table:burstrates}, we first choose a rate in the low range of those values, 30~pulses~hr$^{-1}$. We then find that for a 2-min observation, the probability of detecting zero pulses is about 37\%. 
This implies that for every RRAT present in the surveyed area, there is only a 63\% chance of detection, and we may thus apply a correction factor of $1/0.63=1.6$ to the number of detected RRATs to estimate the true number of RRATs in the surveyed region. 
This conclusion is of course highly dependent on the chosen mean burst rate. 
On the other hand, we can use the average observed burst rate from our sources, 
which is 86~pulses~hr$^{-1}$ (from follow-up observations). In this case, the probability of detecting zero pulses in a 2-min observation is $5.7\%$, yielding a correction factor of 1.06 to the expected number of RRATs in the region. This can thus be taken as a lower limit on the true correction factor that should be applied. Another extreme for the assumed RRAT burst rate can be obtained from the average burst rate of the Parkes Multi-beam Pulsar Survey \citep[PMPS;][]{kle+10} RRATs, which had longer observations and whose rates were thus better determined. From the RRATalog, this average burst rate is 4.2~pulses~hr$^{-1}$. Using this rate in the above analysis, we find that the probability of non-detection in a 2-min observation is 87\%, yielding a correction factor of 7.7 to the number of RRATs in our survey region.

We can also compute the minimum burst rate required for a RRAT to be detected with a given probability. For example, if we set the probability of detection to 99\%, we find that the minimum burst rate required is 138~pulses~hr$^{-1}$ for a 2-min observation. This value is indicated on the burst rate distribution of our new RRATs in Figure~\ref{fig:BRhist}. We note that there are many detected sources below this cutoff, which is statistically unlikely and suggests that there is a large population of sources with low burst rates yet to be discovered in our survey region.

\subsection{Periods}
\label{subsec:discussion period}
We now examine potential biases influencing the observed spin period distributions of detected pulsars and RRATs. 
The first point to address is the potential ambiguity in measured periods. 
While the period of a pulsar found in a periodicity search is easily established, for RRATs there is initially some ambiguity in the inferred period due to the way in which it is computed. That is, one may argue that the computed period is an integer multiple of the true spin period, and that the long periods observed for RRATs are due to this incorrect measurement.
However, as a RRAT is monitored over time, the probability of this occurring diminishes, since a period can be established with high confidence as more pulses are detected. 

One might also expect that a RRAT with a short period will have more chances of detection over a given observation compared to a slower source, since it completes more rotations in the same amount of time. However, it is also possible that this source will instead be detected as a regular pulsar through a periodicity search if its burst rate is sufficiently high. 
For example, some of the RRATs found in our 2-minute survey integrations may be detected in the Fast Fourier Transforms (FFTs) of longer observations, as was sometimes the case in our $\sim20$-minute follow-up observations. 
The classification of a source as a RRAT or normal pulsar is therefore related to the number of pulses detected, which is a function of period, burst rate, and integration time. % 1/N<g<A/sqrt(N), where N is num periods (not num pulses), combined with g it reflects the number of pulses, so this statement is ok.
\cite{mc03} quantify this effect by comparing the significance of detection through a FFT to that of single pulses, for any source with a given period and burst rate. They derive an analytic expression for the ratio of single-pulse to FFT detection significance, $(\mathrm{S/N})_{\mathrm{SP}}/(\mathrm{S/N})_{\mathrm{FFT}}$. %given by Equation~A9 of \cite{mc03}. 
Moreover, they plot \citep[][Figure 13]{mc03} the regimes in which a single-pulse search is more sensitive than a FFT, which indeed shows that as a source emits more pulses, a FFT becomes more effective at detecting it. 

Similarly, \cite{kea10} uses this formalism to deduce the regions of period/nulling fraction space 
in which single-pulse searches are more effective than FFT searches, and thus discoveries would likely be called RRATs. Using these constraints, he then finds that the average observed RRAT period is greater than the average detected pulsar period.

Due to this bias towards longer periods for RRATs, a statistical comparison of the periods of the detected RRAT and pulsar populations is not necessarily representative of the underlying distributions of these sources. However, we can characterize the \emph{detected} populations and test whether those samples of periods are drawn from the same population.

We use the two-sample Kolmogorov-Smirnov (K-S) test to compare the distributions of the properties mentioned above.  
We add the sample of 20 RRATs\footnote{We use 20 of the 21 RRATs presented here, since we exclude RRAT J0054+69 for which we could not reliably measure the period, as described in the text.} found in this work to the 61 RRATs listed in the RRATalog. We then perform a K-S test between the periods of the RRATs and the $\sim$2000 radio pulsars listed in the ATNF Catalogue (where we have excluded MSPs and magnetars). This K-S test yields a p-value of $6 \times 10^{-14}$, implying that the samples are drawn from extremely different distributions, as expected. 

We can also compare the periods of the new RRATs to those of all known RRATs, listed in the RRATalog. This yields a p-value of 0.16, meaning that the samples are consistent with coming from the same distribution. 
This is not surprising, since the sensitivities of recent surveys to RRATs are comparable to one another (e.g. Pulsar Arecibo L-band Feed Array survey, \citealt{dcm+09}; Parkes intermediate and high latitude surveys, \citealt{bsb10}; PMPS, \citealt{kle+10}).

Finally, we compare the periods of the new RRATs to those of the 128 pulsars discovered in the Drift-scan and GBNCC surveys. This yields a p-value of 0.002, implying that the detected RRAT and pulsar period samples are likely drawn from different distributions. 
We plot the period distributions of the RRATs and pulsars found in each of the Drift-scan and GBNCC surveys in Figure~\ref{fig:DM-Phist} (left). These histograms have been normalized to account for the total number of sources. 

While we find that RRATs have preferentially longer periods than canonical pulsars (i.e. pulsars which were initially detected in a periodicity search), we mention above that this may be related to a classification bias. We can avoid this bias by comparing the periods of single-pulse emitting pulsars to those of pulsars from which single pulses are not observed. Since single-pulse searches have uniform sensitivity regardless of the spin period of a pulsar/RRAT (as long as the pulses are identical), if fast rotators indeed emit bright RRAT-like pulses, we expect to detect them. We plot in Figure~\ref{fig:Pcomparison-sps} the period distribution of pulsars discovered in the GBNCC survey as well as known and newly discovered pulsars detected in the Drift-scan survey. We distinguish between sources from which single pulses were and were not detected. We see that generally, rapidly rotating pulsars do not emit bright single pulses, whereas slower pulsars often do. Moreover, a K-S test between these samples yields a p-value of $10^{-13}$, indicating different distributions.

Again, these distributions may be influenced by detection biases. The S/N of a single-pulse detection depends on the width and amplitude of a given pulse. For a fixed flux density and duty cycle, we find that the flux per pulse is the same regardless of pulsar period, and therefore expect only a mild bias in the distribution of sources detected via single pulses, due to broader pulse widths in slow pulses. On the other hand, if the duty cycle varies with period, for example as $P^{-1/2}$ \citep[e.g.][]{ran93}, then we expect longer-period pulsars to have larger flux per pulse, which would cause some bias in the observed distribution. Although these factors account for some of the difference between the period distributions, the observed effect (Figure~\ref{fig:Pcomparison-sps}) is so pronounced that it could well also be caused by an intrinsic difference between the slow and fast pulsar populations and suggests that bright single-pulse emitters are preferentially slower rotators. Detailed quantitative follow-up with corrections for differences in pulse widths and fluences will allow firm conclusions.

\subsection{Dispersion measures}
\label{subsec:discussion dm}
We now turn to examining the biases affecting the DM distribution of a given survey's discoveries. This property is heavily survey-dependent, since DM varies greatly depending on the surveyed region due to the non-uniform distribution of free electrons throughout the Galaxy. 
For this reason, we compare only the DMs of sources found in the same survey. 

Another effect that must be considered is the dependence of a survey's sensitivity on DM. Since interstellar scattering timescales increase with DM and with decreasing observing frequency (as $f^{-4.4}$, e.g. \citealt{ric90}), 
surveys are typically limited in their sensitivity to short-period pulsars at high DMs. This effect is mostly uniform for pulsars and RRATs, if their intrinsic period distributions are the same, although the narrower pulses of RRATs would likely be affected more by the diminished S/N due to scattering. 

\begin{figure*}
\centering
\includegraphics[width=0.65\linewidth]{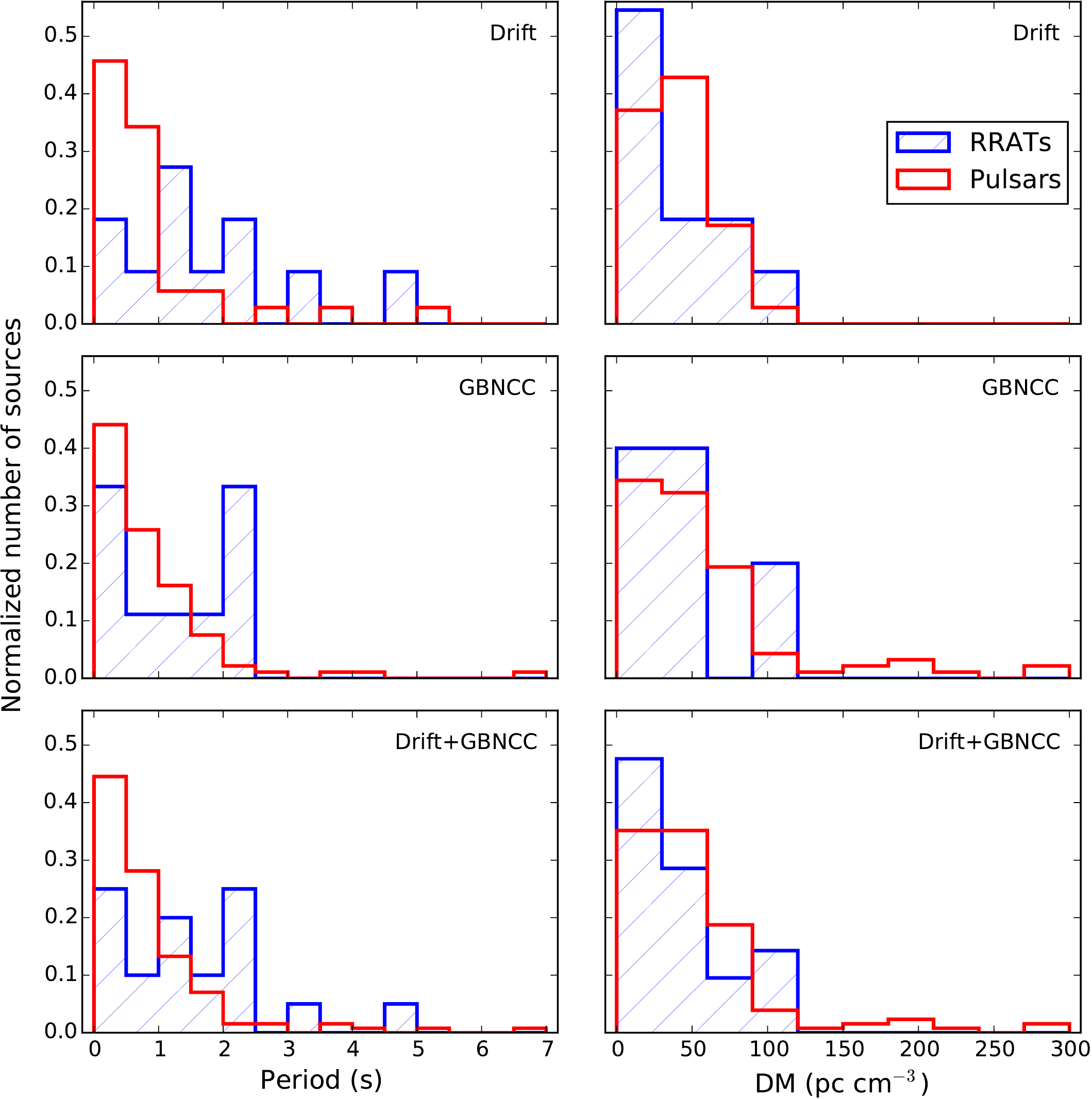}
\caption{Comparison of period (left) and DM (right) distributions of RRATs (blue-hatched) and pulsars (red) found in the Drift-scan and GBNCC surveys. The histograms have been normalized and include 11 Drift-scan and 10 GBNCC RRATs, and 35 Drift-scan and 93 GBNCC pulsars. As described in Sections~\ref{subsec:discussion period} and~\ref{subsec:discussion dm}, the observed period distributions of pulsars and RRATs differ, whereas their DM distributions are similar.}
\label{fig:DM-Phist}
\end{figure*}

\begin{figure}
    \centering
    \includegraphics[width=\linewidth]{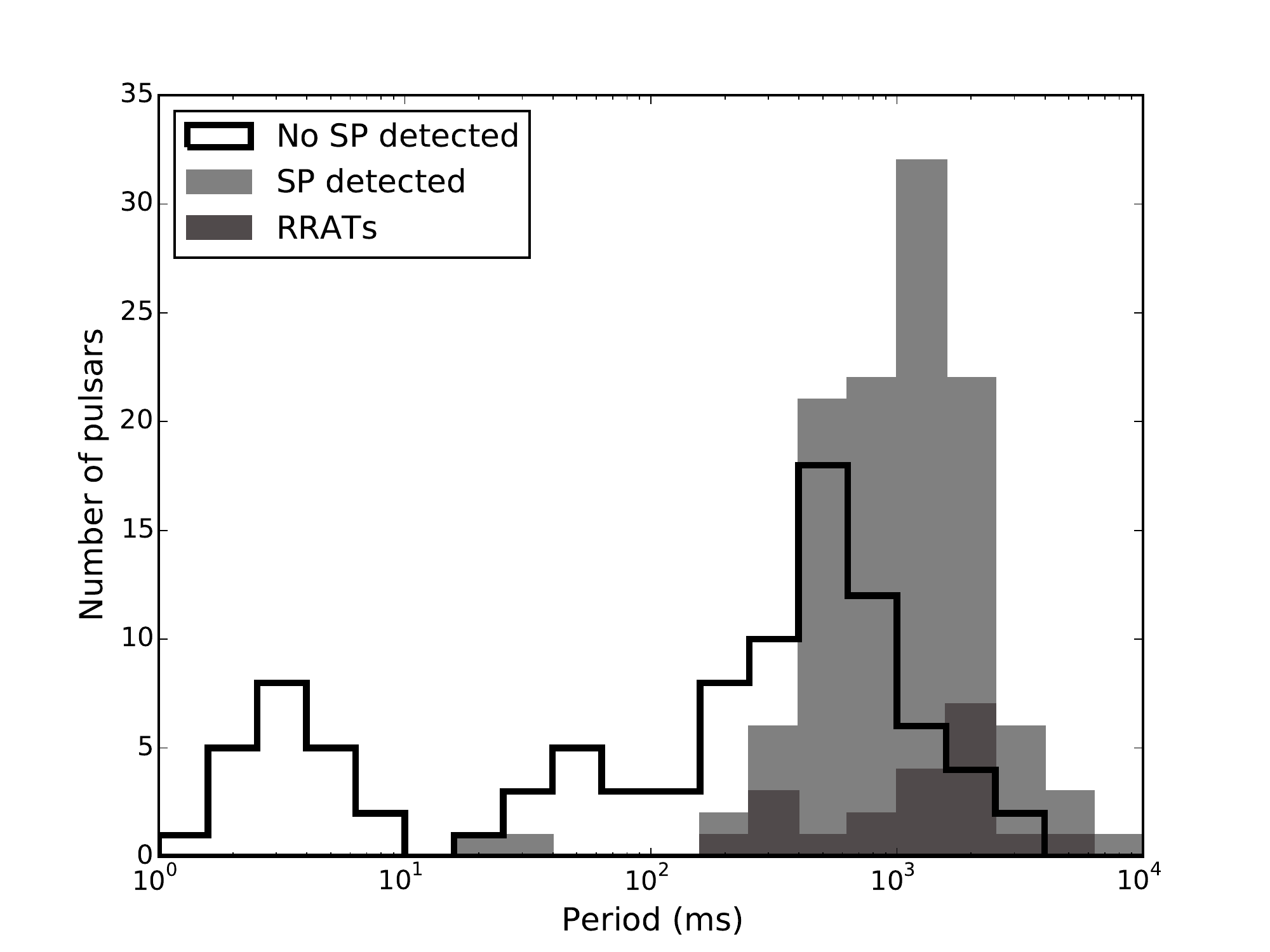}
    \caption{Period distribution for GBNCC discoveries and Drift-scan discoveries and detections of known pulsars. Sources detectable through single pulses are plotted in gray, with RRATs highlighted in dark gray. Sources not detectable through single pulses are outlined in black. All sources except the RRATs were detected in a periodicity search.\label{fig:Pcomparison-sps}}
\end{figure}

The DM distributions of RRATs and pulsars discovered in each of the Drift-scan and GBNCC surveys are shown in Figure~\ref{fig:DM-Phist} (right). Note that these histograms have been normalized to account for the total number of sources. 
Performing a 2-sample K-S test between the DMs of the RRATs and pulsars in the Drift-scan survey, we obtain a p-value of 0.68, indicating that the samples are consistent with coming from the same DM distribution. Repeating this for the sources in the GBNCC survey, we obtain a p-value of 0.97, again implying a shared underlying DM distribution. 

\subsection{Spatial distributions}
\label{subsec:discussion spatial}
We can similarly compare the spatial distributions of RRATs and pulsars. Since different surveys cover different areas of the sky, meaningful comparisons are again only between sources found in the same survey.

We compare the Galactic longitudes ($l$) and latitudes ($b$) of pulsars to those of RRATs, for the Drift-scan and GBNCC surveys. A K-S test for the $l$ and $b$ distributions in the Drift-scan survey yields p-values of 0.88 for both, implying highly consistent spatial distributions. The same test for the GBNCC sources yields p-values of 0.31 ($l$) and 0.40 ($b$), again implying that the two samples are consistent with having a shared spatial distribution.

Examining the spatial distributions of sources is of interest since the Galactic position of a source may help constrain its age. For example, we expect young pulsars to be within the Galactic plane, where they formed, while older pulsars are found outside of the plane. The above results thus suggest that, according to spatial distributions at least, RRATs and pulsars are consistent with having similar underlying properties.

\subsection{Fast radio bursts}
We did not find any FRBs in either the Drift-scan or the GBNCC survey. Since there are only several FRBs known \citep{tsb+13,sch+14,bsb14}, their DM distribution is not well constrained. Moreover, the published FRB rates have all been inferred from results of surveys conducted at a frequency of 1.4~GHz, making it difficult to predict a rate for surveys at 350~MHz since the spectral index for FRBs is not known.

Assuming FRBs to have a uniform distribution of DMs, we can use the total observing time of each survey and telescope beam size to obtain an upper limit on the FRB rate, given that none have been detected. 
For the Drift-scan, assuming a Poisson process, we find a 99\% upper limit on the FRB rate of $\sim 1\times10^4~\mathrm{sky}^{-1}~\mathrm{day}^{-1}$ above the minimum flux density, $S_\mathrm{min}=260$~mJy. 
%The beam size of the GBT at 350~MHz is 0.28 deg$^2$. For the Drift-scan then, the upper limit on FRB rate is $1/(1491~\mathrm{hr} \times 0.28~\mathrm{deg}^2) \times 41253~\mathrm{deg}^2/\mathrm{sky} \sim 2\times10^3~\mathrm{sky}^{-1}~\mathrm{day}^{-1}$.
% For 99% confidence upper lim, multiply rate by 4.6 (P(0)=e^-a*a^0/0! = e^-a = 0.01 --> a=-ln(0.01) = 4.6
For comparison, the rate inferred by \cite{tsb+13} is $1\times10^4~\mathrm{sky}^{-1}~\mathrm{day}^{-1}$, whereas the rate inferred by \cite{bsb14} is $2\times10^3~\mathrm{sky}^{-1}~\mathrm{day}^{-1}$, both for a fluence of $\sim3$~Jy~s. 
For much of the GBNCC data analyzed with {\tt RRATtrap}, the maximum DM searched was 500~\dmunit, which is typically lower than the DMs of FRBs observed thus far \citep[e.g.][]{tsb+13}. We therefore do not quote here an event rate limit based on the GBNCC survey. 
A more thorough analysis will be presented in the future, placing better constraints on the FRB rates based on these surveys.

\section{Conclusions}
\label{sec:conclusion}
We have developed and used a new single-pulse sifting algorithm, {\tt RRATtrap}, in order to find RRATs in GBT Drift-scan and GBNCC pulsar survey data, resulting in the discovery of 21 RRATs. We have characterized these RRATs by refining their periods, DMs, and burst rates. 
The new RRATs have DMs ranging from 15 to 97~\dmunit, periods of 240~ms to 3.4~s, and burst rates of 20 to 400~pulses~hr$^{-1}$ at 350~MHz. 
We have also obtained phase-coherent timing solutions for four of the new RRATs.

We have presented our new RRAT discoveries in the context of the known RRAT population, as well as the radio pulsar population. We described the various detection biases that exist between RRATs and pulsars, and took note of these when conducting statistical comparisons of the two populations' properties. 
We found that the DM and spatial distributions of RRATs and pulsars discovered in the same survey are consistent with having a shared underlying distribution. 
In contrast, the period distribution of detected RRATs is significantly different than that of detected canonical pulsars. 
This may be an effect of the classification biases related to period and burst rate values of RRATs, and is not necessarily representative of the underlying RRAT and pulsar populations. 
We also find evidence that slower pulsars (i.e. $P>200$~ms) are more likely to emit RRAT-like bright single pulses than are faster pulsars ($P<200$~ms), although this conclusion is tentative.

We have also shown that RRATs have a distribution of burst rates which can affect their classification as single-pulse or periodicity sources and also bears consequences for their discovery in surveys. 
We note that while we observed a certain distribution of burst rates, the underlying distribution is not known due to heavy observational biases against detecting RRATs with low burst rates. However, the increasing diversity in 
properties such as intermittency timescales of transients
has begun to hint at a continuum of behavior, supporting the idea that RRATs are related to nulling pulsars and may simply represent extreme cases of the latter. 

\section{Acknowledgments}
\label{sec:acks}
The National Radio Astronomy Observatory is a facility of the National Science Foundation operated under cooperative agreement by Associated Universities, Inc. 
LOFAR, the Low Frequency Array designed and constructed by ASTRON, has facilities in several countries, that are owned by various parties (each with their own funding sources), and that are collectively operated by the International LOFAR Telescope (ILT) foundation under a joint scientific policy. 
We thank the referee for helpful comments. 
We thank Compute Canada and the McGill Center for High Performance Computing and Calcul Quebec for provision and maintenance of the Guillimin supercomputer and related resources. 
V.M.K. acknowledges support from an NSERC Discovery Grant and Accelerator Supplement, the FRQNT Centre de Recherche en Astrophysique du Qu\'ebec, an R. Howard Webster Foundation Fellowship from the Canadian Institute for Advanced Research (CIFAR), the Canada Research Chairs Program and the Lorne Trottier Chair in Astrophysics and Cosmology. 
J.W.T.H. acknowledges funding from an NWO Vidi fellowship and ERC Starting Grant ``DRAGNET" (337062). 
I.H.S. was supported by an NSERC Discovery Grant, by CIFAR, and by the Canada Foundation for Innovation.

\clearpage


\begin{thebibliography}{}
\expandafter\ifx\csname natexlab\endcsname\relax\def\natexlab#1{#1}\fi

\bibitem[{{Archibald} {et~al.}(2009){Archibald}, {Stairs}, {Ransom}, {Kaspi},
  {Kondratiev}, {Lorimer}, {McLaughlin}, {Boyles}, {Hessels}, {Lynch}, {van
  Leeuwen}, {Roberts}, {Jenet}, {Champion}, {Rosen}, {Barlow}, {Dunlap}, \&
  {Remillard}}]{asr+09}
{Archibald}, A.~M., {Stairs}, I.~H., {Ransom}, S.~M., {et~al.} 2009, Science,
  324, 1411

\bibitem[{{Arts} {et~al.}(2013){Arts}, {Kant}, \& {Wijnholds}}]{akw13}
{Arts}, M.~J., {Kant}, G.~W., \& {Wijnholds}, S.~J. 2013, Sensitivity
  Approximations for Regular Aperture Arrays, Internal Memo SKA-ASTRON-RP-473,
  ASTRON, The Netherlands

\bibitem[{{Bandura} {et~al.}(2014){Bandura}, {Addison}, {Amiri}, {Bond},
  {Campbell-Wilson}, {Connor}, {Cliche}, {Davis}, {Deng}, {Denman}, {Dobbs},
  {Fandino}, {Gibbs}, {Gilbert}, {Halpern}, {Hanna}, {Hincks}, {Hinshaw},
  {H{\"o}fer}, {Klages}, {Landecker}, {Masui}, {Mena Parra}, {Newburgh}, {Pen},
  {Peterson}, {Recnik}, {Shaw}, {Sigurdson}, {Sitwell}, {Smecher}, {Smegal},
  {Vanderlinde}, \& {Wiebe}}]{baa+14}
{Bandura}, K., {Addison}, G.~E., {Amiri}, M., {et~al.} 2014, in Society of
  Photo-Optical Instrumentation Engineers (SPIE) Conference Series, Vol. 9145,
  Society of Photo-Optical Instrumentation Engineers (SPIE) Conference Series,
  22

\bibitem[{{Boyles} {et~al.}(2013){Boyles}, {Lynch}, {Ransom}, {Stairs},
  {Lorimer}, {McLaughlin}, {Hessels}, {Kaspi}, {Kondratiev}, {Archibald},
  {Berndsen}, {Cardoso}, {Cherry}, {Epstein}, {Karako-Argaman}, {McPhee},
  {Pennucci}, {Roberts}, {Stovall}, \& {van Leeuwen}}]{blr+13}
{Boyles}, J., {Lynch}, R.~S., {Ransom}, S.~M., {et~al.} 2013, \apj, 763, 80

\bibitem[{{Burke-Spolaor} \& {Bailes}(2010)}]{bsb10}
{Burke-Spolaor}, S., \& {Bailes}, M. 2010, \mnras, 402, 855

\bibitem[{{Burke-Spolaor} {et~al.}(2011){Burke-Spolaor}, {Bailes}, {Ekers},
  {Macquart}, \& {Crawford}}]{bbe+11}
{Burke-Spolaor}, S., {Bailes}, M., {Ekers}, R., {Macquart}, J.-P., \&
  {Crawford}, III, F. 2011, \apj, 727, 18

\bibitem[{{Burke-Spolaor} \& {Bannister}(2014)}]{bsb14}
{Burke-Spolaor}, S., \& {Bannister}, K.~W. 2014, \apj, 792, 19

\bibitem[{{Burke-Spolaor} {et~al.}(2012){Burke-Spolaor}, {Johnston}, {Bailes},
  {Bates}, {Bhat}, {Burgay}, {Champion}, {D'Amico}, {Keith}, {Kramer}, {Levin},
  {Milia}, {Possenti}, {Stappers}, \& {van Straten}}]{bjs+12}
{Burke-Spolaor}, S., {Johnston}, S., {Bailes}, M., {et~al.} 2012, \mnras, 423,
  1351

\bibitem[{{Chen} \& {Ruderman}(1993)}]{cr93}
{Chen}, K., \& {Ruderman}, M. 1993, \apj, 402, 264

\bibitem[{{Cordes}(2002)}]{cor02}
{Cordes}, J.~M. 2002, in Astronomical Society of the Pacific Conference Series,
  Vol. 278, Single-Dish Radio Astronomy: Techniques and Applications, ed.
  S.~{Stanimirovic}, D.~{Altschuler}, P.~{Goldsmith}, \& C.~{Salter}, 227--250

\bibitem[{{Cordes} \& {Lazio}(2002)}]{cl02}
{Cordes}, J.~M., \& {Lazio}, T.~J.~W. 2002, ArXiv Astrophysics e-prints,
  arXiv:astro-ph/0207156

\bibitem[{{Cordes} \& {McLaughlin}(2003)}]{cm03}
{Cordes}, J.~M., \& {McLaughlin}, M.~A. 2003, \apj, 596, 1142

\bibitem[{{Cordes} \& {Shannon}(2008)}]{cs08}
{Cordes}, J.~M., \& {Shannon}, R.~M. 2008, \apj, 682, 1152

\bibitem[{{Deneva} {et~al.}(2009){Deneva}, {Cordes}, {McLaughlin}, {Nice},
  {Lorimer}, {Crawford}, {Bhat}, {Camilo}, {Champion}, {Freire}, {Edel},
  {Kondratiev}, {Hessels}, {Jenet}, {Kasian}, {Kaspi}, {Kramer}, {Lazarus},
  {Ransom}, {Stairs}, {Stappers}, {van Leeuwen}, {Brazier}, {Venkataraman},
  {Zollweg}, \& {Bogdanov}}]{dcm+09}
{Deneva}, J.~S., {Cordes}, J.~M., {McLaughlin}, M.~A., {et~al.} 2009, \apj,
  703, 2259

\bibitem[{{DuPlain} {et~al.}(2008){DuPlain}, {Ransom}, {Demorest}, {Brandt},
  {Ford}, \& {Shelton}}]{drd+08}
{DuPlain}, R., {Ransom}, S., {Demorest}, P., {et~al.} 2008, in Society of
  Photo-Optical Instrumentation Engineers (SPIE) Conference Series, Vol. 7019,
  Society of Photo-Optical Instrumentation Engineers (SPIE) Conference Series

\bibitem[{{Hobbs} {et~al.}(2010){Hobbs}, {Archibald}, {Arzoumanian}, {Backer},
  {Bailes}, {Bhat}, {Burgay}, {Burke-Spolaor}, {Champion}, {Cognard}, {Coles},
  {Cordes}, {Demorest}, {Desvignes}, {Ferdman}, {Finn}, {Freire}, {Gonzalez},
  {Hessels}, {Hotan}, {Janssen}, {Jenet}, {Jessner}, {Jordan}, {Kaspi},
  {Kramer}, {Kondratiev}, {Lazio}, {Lazaridis}, {Lee}, {Levin}, {Lommen},
  {Lorimer}, {Lynch}, {Lyne}, {Manchester}, {McLaughlin}, {Nice}, {Oslowski},
  {Pilia}, {Possenti}, {Purver}, {Ransom}, {Reynolds}, {Sanidas}, {Sarkissian},
  {Sesana}, {Shannon}, {Siemens}, {Stairs}, {Stappers}, {Stinebring},
  {Theureau}, {van Haasteren}, {van Straten}, {Verbiest}, {Yardley}, \&
  {You}}]{haa+10}
{Hobbs}, G., {Archibald}, A., {Arzoumanian}, Z., {et~al.} 2010, Classical and
  Quantum Gravity, 27, 084013

\bibitem[{{Hotan} {et~al.}(2004){Hotan}, {van Straten}, \&
  {Manchester}}]{hsm04}
{Hotan}, A.~W., {van Straten}, W., \& {Manchester}, R.~N. 2004, \pasa, 21, 302

\bibitem[{{Kaplan} {et~al.}(2012){Kaplan}, {Stovall}, {Ransom}, {Roberts},
  {Kotulla}, {Archibald}, {Biwer}, {Boyles}, {Dartez}, {Day}, {Ford}, {Garcia},
  {Hessels}, {Jenet}, {Karako}, {Kaspi}, {Kondratiev}, {Lorimer}, {Lynch},
  {McLaughlin}, {Rohr}, {Siemens}, {Stairs}, \& {van Leeuwen}}]{ksr+12}
{Kaplan}, D.~L., {Stovall}, K., {Ransom}, S.~M., {et~al.} 2012, \apj, 753, 174

\bibitem[{{Keane}(2010)}]{kea10}
{Keane}, E.~F. 2010, PhD thesis, University of Manchester

\bibitem[{{Keane} \& {Kramer}(2008)}]{kk08}
{Keane}, E.~F., \& {Kramer}, M. 2008, \mnras, 391, 2009

\bibitem[{{Keane} {et~al.}(2010){Keane}, {Ludovici}, {Eatough}, {Kramer},
  {Lyne}, {McLaughlin}, \& {Stappers}}]{kle+10}
{Keane}, E.~F., {Ludovici}, D.~A., {Eatough}, R.~P., {et~al.} 2010, \mnras,
  401, 1057

\bibitem[{{Keane} \& {Petroff}(2015)}]{kp15}
{Keane}, E.~F., \& {Petroff}, E. 2015, \mnras, 447, 2852

\bibitem[{{Kulkarni} {et~al.}(2014){Kulkarni}, {Ofek}, {Neill}, {Zheng}, \&
  {Juric}}]{kon+14}
{Kulkarni}, S.~R., {Ofek}, E.~O., {Neill}, J.~D., {Zheng}, Z., \& {Juric}, M.
  2014, \apj, 797, 70

\bibitem[{{Li}(2006)}]{li06}
{Li}, X.-D. 2006, \apjl, 646, L139

\bibitem[{{Lorimer} {et~al.}(2007){Lorimer}, {Bailes}, {McLaughlin},
  {Narkevic}, \& {Crawford}}]{lbm+07}
{Lorimer}, D.~R., {Bailes}, M., {McLaughlin}, M.~A., {Narkevic}, D.~J., \&
  {Crawford}, F. 2007, Science, 318, 777

\bibitem[{{Lorimer} \& {Kramer}(2004)}]{lk04}
{Lorimer}, D.~R., \& {Kramer}, M. 2004, {\emph{Handbook of Pulsar Astronomy}}
  (Cambridge University Press)

\bibitem[{{Lynch} {et~al.}(2013){Lynch}, {Boyles}, {Ransom}, {Stairs},
  {Lorimer}, {McLaughlin}, {Hessels}, {Kaspi}, {Kondratiev}, {Archibald},
  {Berndsen}, {Cardoso}, {Cherry}, {Epstein}, {Karako-Argaman}, {McPhee},
  {Pennucci}, {Roberts}, {Stovall}, \& {van Leeuwen}}]{lbr+13}
{Lynch}, R.~S., {Boyles}, J., {Ransom}, S.~M., {et~al.} 2013, \apj, 763, 81

\bibitem[{{Lyne} {et~al.}(2009){Lyne}, {McLaughlin}, {Keane}, {Kramer},
  {Espinoza}, {Stappers}, {Palliyaguru}, \& {Miller}}]{lmk+09}
{Lyne}, A.~G., {McLaughlin}, M.~A., {Keane}, E.~F., {et~al.} 2009, \mnras, 400,
  1439

\bibitem[{{Manchester} {et~al.}(2005){Manchester}, {Hobbs}, {Teoh}, \&
  {Hobbs}}]{mht+05}
{Manchester}, R.~N., {Hobbs}, G.~B., {Teoh}, A., \& {Hobbs}, M. 2005, \aj, 129,
  1993

\bibitem[{{McLaughlin} \& {Cordes}(2003)}]{mc03}
{McLaughlin}, M.~A., \& {Cordes}, J.~M. 2003, \apj, 596, 982

\bibitem[{{McLaughlin} {et~al.}(2006){McLaughlin}, {Lyne}, {Lorimer}, {Kramer},
  {Faulkner}, {Manchester}, {Cordes}, {Camilo}, {Possenti}, {Stairs}, {Hobbs},
  {D'Amico}, {Burgay}, \& {O'Brien}}]{mll+06}
{McLaughlin}, M.~A., {Lyne}, A.~G., {Lorimer}, D.~R., {et~al.} 2006, \nat, 439,
  817

\bibitem[{{Olausen} \& {Kaspi}(2014)}]{ok14}
{Olausen}, S.~A., \& {Kaspi}, V.~M. 2014, \apjs, 212, 6

\bibitem[{{Popov} {et~al.}(2006){Popov}, {Turolla}, \& {Possenti}}]{ptp06}
{Popov}, S.~B., {Turolla}, R., \& {Possenti}, A. 2006, \mnras, 369, L23

\bibitem[{{Rankin}(1993)}]{ran93}
{Rankin}, J.~M. 1993, \apj, 405, 285

\bibitem[{{Ransom}(2001)}]{ran01}
{Ransom}, S.~M. 2001, PhD thesis, Harvard University

\bibitem[{{Ransom} {et~al.}(2014){Ransom}, {Stairs}, {Archibald}, {Hessels},
  {Kaplan}, {van Kerkwijk}, {Boyles}, {Deller}, {Chatterjee},
  {Schechtman-Rook}, {Berndsen}, {Lynch}, {Lorimer}, {Karako-Argaman}, {Kaspi},
  {Kondratiev}, {McLaughlin}, {van Leeuwen}, {Rosen}, {Roberts}, \&
  {Stovall}}]{rsa+14}
{Ransom}, S.~M., {Stairs}, I.~H., {Archibald}, A.~M., {et~al.} 2014, \nat, 505,
  520

\bibitem[{{Rickett}(1990)}]{ric90}
{Rickett}, B.~J. 1990, \araa, 28, 561

\bibitem[{{Ritchings}(1976)}]{rit76}
{Ritchings}, R.~T. 1976, \mnras, 176, 249

\bibitem[{{Spitler} {et~al.}(2014){Spitler}, {Cordes}, {Hessels}, {Lorimer},
  {McLaughlin}, {Chatterjee}, {Crawford}, {Deneva}, {Kaspi}, {Wharton},
  {Allen}, {Bogdanov}, {Brazier}, {Camilo}, {Freire}, {Jenet},
  {Karako-Argaman}, {Knispel}, {Lazarus}, {Lee}, {van Leeuwen}, {Lynch},
  {Ransom}, {Scholz}, {Siemens}, {Stairs}, {Stovall}, {Swiggum},
  {Venkataraman}, {Zhu}, {Aulbert}, \& {Fehrmann}}]{sch+14}
{Spitler}, L.~G., {Cordes}, J.~M., {Hessels}, J.~W.~T., {et~al.} 2014, \apj,
  790, 101

\bibitem[{{Stappers} {et~al.}(2011){Stappers}, {Hessels}, {Alexov}, {Anderson},
  {Coenen}, {Hassall}, {Karastergiou}, {Kondratiev}, {Kramer}, {van Leeuwen},
  {Mol}, {Noutsos}, {Romein}, {Weltevrede}, {Fender}, {Wijers}, {B{\"a}hren},
  {Bell}, {Broderick}, {Daw}, {Dhillon}, {Eisl{\"o}ffel}, {Falcke},
  {Griessmeier}, {Law}, {Markoff}, {Miller-Jones}, {Scheers}, {Spreeuw},
  {Swinbank}, {Ter Veen}, {Wise}, {Wucknitz}, {Zarka}, {Anderson}, {Asgekar},
  {Avruch}, {Beck}, {Bennema}, {Bentum}, {Best}, {Bregman}, {Brentjens}, {van
  de Brink}, {Broekema}, {Brouw}, {Br{\"u}ggen}, {de Bruyn}, {Butcher},
  {Ciardi}, {Conway}, {Dettmar}, {van Duin}, {van Enst}, {Garrett}, {Gerbers},
  {Grit}, {Gunst}, {van Haarlem}, {Hamaker}, {Heald}, {Hoeft}, {Holties},
  {Horneffer}, {Koopmans}, {Kuper}, {Loose}, {Maat}, {McKay-Bukowski},
  {McKean}, {Miley}, {Morganti}, {Nijboer}, {Noordam}, {Norden}, {Olofsson},
  {Pandey-Pommier}, {Polatidis}, {Reich}, {R{\"o}ttgering}, {Schoenmakers},
  {Sluman}, {Smirnov}, {Steinmetz}, {Sterks}, {Tagger}, {Tang}, {Vermeulen},
  {Vermaas}, {Vogt}, {de Vos}, {Wijnholds}, {Yatawatta}, \& {Zensus}}]{sha+11}
{Stappers}, B.~W., {Hessels}, J.~W.~T., {Alexov}, A., {et~al.} 2011, \aap, 530,
  A80

\bibitem[{{Stovall} {et~al.}(2014){Stovall}, {Lynch}, {Ransom}, {Archibald},
  {Banaszak}, {Biwer}, {Boyles}, {Dartez}, {Day}, {Ford}, {Flanigan}, {Garcia},
  {Hessels}, {Hinojosa}, {Jenet}, {Kaplan}, {Karako-Argaman}, {Kaspi},
  {Kondratiev}, {Leake}, {Lorimer}, {Lunsford}, {Martinez}, {Mata},
  {McLaughlin}, {Roberts}, {Rohr}, {Siemens}, {Stairs}, {van Leeuwen},
  {Walker}, \& {Wells}}]{slr+14}
{Stovall}, K., {Lynch}, R.~S., {Ransom}, S.~M., {et~al.} 2014, \apj, 791, 67

\bibitem[{{Taylor}(1992)}]{tay92}
{Taylor}, J.~H. 1992, Royal Society of London Philosophical Transactions Series
  A, 341, 117

\bibitem[{{Thornton} {et~al.}(2013){Thornton}, {Stappers}, {Bailes},
  {Barsdell}, {Bates}, {Bhat}, {Burgay}, {Burke-Spolaor}, {Champion}, {Coster},
  {D'Amico}, {Jameson}, {Johnston}, {Keith}, {Kramer}, {Levin}, {Milia}, {Ng},
  {Possenti}, \& {van Straten}}]{tsb+13}
{Thornton}, D., {Stappers}, B., {Bailes}, M., {et~al.} 2013, Science, 341, 53

\bibitem[{{van Haarlem} {et~al.}(2013){van Haarlem}, {Wise}, {Gunst}, {Heald},
  {McKean}, {Hessels}, {de Bruyn}, {Nijboer}, {Swinbank}, {Fallows},
  {Brentjens}, {Nelles}, {Beck}, {Falcke}, {Fender}, {H{\"o}randel},
  {Koopmans}, {Mann}, {Miley}, {R{\"o}ttgering}, {Stappers}, {Wijers},
  {Zaroubi}, {van den Akker}, {Alexov}, {Anderson}, {Anderson}, {van Ardenne},
  {Arts}, {Asgekar}, {Avruch}, {Batejat}, {B{\"a}hren}, {Bell}, {Bell}, {van
  Bemmel}, {Bennema}, {Bentum}, {Bernardi}, {Best}, {B{\^i}rzan}, {Bonafede},
  {Boonstra}, {Braun}, {Bregman}, {Breitling}, {van de Brink}, {Broderick},
  {Broekema}, {Brouw}, {Br{\"u}ggen}, {Butcher}, {van Cappellen}, {Ciardi},
  {Coenen}, {Conway}, {Coolen}, {Corstanje}, {Damstra}, {Davies}, {Deller},
  {Dettmar}, {van Diepen}, {Dijkstra}, {Donker}, {Doorduin}, {Dromer}, {Drost},
  {van Duin}, {Eisl{\"o}ffel}, {van Enst}, {Ferrari}, {Frieswijk}, {Gankema},
  {Garrett}, {de Gasperin}, {Gerbers}, {de Geus}, {Grie{\ss}meier}, {Grit},
  {Gruppen}, {Hamaker}, {Hassall}, {Hoeft}, {Holties}, {Horneffer}, {van der
  Horst}, {van Houwelingen}, {Huijgen}, {Iacobelli}, {Intema}, {Jackson},
  {Jelic}, {de Jong}, {Juette}, {Kant}, {Karastergiou}, {Koers}, {Kollen},
  {Kondratiev}, {Kooistra}, {Koopman}, {Koster}, {Kuniyoshi}, {Kramer},
  {Kuper}, {Lambropoulos}, {Law}, {van Leeuwen}, {Lemaitre}, {Loose}, {Maat},
  {Macario}, {Markoff}, {Masters}, {McFadden}, {McKay-Bukowski}, {Meijering},
  {Meulman}, {Mevius}, {Middelberg}, {Millenaar}, {Miller-Jones}, {Mohan},
  {Mol}, {Morawietz}, {Morganti}, {Mulcahy}, {Mulder}, {Munk}, {Nieuwenhuis},
  {van Nieuwpoort}, {Noordam}, {Norden}, {Noutsos}, {Offringa}, {Olofsson},
  {Omar}, {Orr{\'u}}, {Overeem}, {Paas}, {Pandey-Pommier}, {Pandey}, {Pizzo},
  {Polatidis}, {Rafferty}, {Rawlings}, {Reich}, {de Reijer}, {Reitsma},
  {Renting}, {Riemers}, {Rol}, {Romein}, {Roosjen}, {Ruiter}, {Scaife}, {van
  der Schaaf}, {Scheers}, {Schellart}, {Schoenmakers}, {Schoonderbeek},
  {Serylak}, {Shulevski}, {Sluman}, {Smirnov}, {Sobey}, {Spreeuw}, {Steinmetz},
  {Sterks}, {Stiepel}, {Stuurwold}, {Tagger}, {Tang}, {Tasse}, {Thomas},
  {Thoudam}, {Toribio}, {van der Tol}, {Usov}, {van Veelen}, {van der Veen},
  {ter Veen}, {Verbiest}, {Vermeulen}, {Vermaas}, {Vocks}, {Vogt}, {de Vos},
  {van der Wal}, {van Weeren}, {Weggemans}, {Weltevrede}, {White}, {Wijnholds},
  {Wilhelmsson}, {Wucknitz}, {Yatawatta}, {Zarka}, {Zensus}, \& {van
  Zwieten}}]{hwg+13}
{van Haarlem}, M.~P., {Wise}, M.~W., {Gunst}, A.~W., {et~al.} 2013, \aap, 556,
  A2

\bibitem[{{van Straten} \& {Bailes}(2011)}]{sb11}
{van Straten}, W., \& {Bailes}, M. 2011, \pasa, 28, 1

\bibitem[{{Wang} {et~al.}(2007){Wang}, {Manchester}, \& {Johnston}}]{wmj07}
{Wang}, N., {Manchester}, R.~N., \& {Johnston}, S. 2007, \mnras, 377, 1383

\bibitem[{{Weltevrede} {et~al.}(2006){Weltevrede}, {Stappers}, {Rankin}, \&
  {Wright}}]{wsr+06}
{Weltevrede}, P., {Stappers}, B.~W., {Rankin}, J.~M., \& {Wright}, G.~A.~E.
  2006, \apjl, 645, L149

\bibitem[{{Zhang} {et~al.}(2007){Zhang}, {Gil}, \& {Dyks}}]{zgd07}
{Zhang}, B., {Gil}, J., \& {Dyks}, J. 2007, \mnras, 374, 1103

\end{thebibliography}
\end{document}